\documentclass{emulateapj}
\usepackage{natbib,color,ulem,comment}

\newcommand{\K}{\,\mathrm{K}}

\newcommand{\x}[1]{}

\newcommand{\y}{\color{blue}}

\newcommand{\bd}{}
\newcommand{\bdd}{}

\def\refnew#1{(\ref{#1})}

\shorttitle{The ``boil-off'' of low-mass planets}
\shortauthors{Owen, J. E. \& Wu, Y.}

\begin{document}


\title{Atmospheres of low-mass planets: the ``boil-off''}


\author{James E. Owen\altaffilmark{1}}
\affil{Institute for Advanced Study, Einstein Drive, Princeton NJ, 08540, USA}
\email{jowen@ias.edu}
\and
\author{Yanqin Wu}
\affil{Department of Astronomy and Astrophysics, University of Toronto, Toronto, ON M5S 3H4, CANADA}
%
%


\altaffiltext{1}{Hubble Fellow}


\begin{abstract} We show that, for a low-mass planet that orbits its host star within a few tenths of an AU (like the majority of the {\it Kepler} planets), the atmosphere it was able to accumulate while embedded in the proto-planetary disk may not survive unscathed after the disk disperses. This gas envelope, if more massive than a few percent of the core (with a mass below $10 M_\oplus$), has a cooling time that is much longer than the time-scale on which the planet exits the disk. As such, it could not have contracted significantly from its original size, of order the Bondi radius.  So a newly exposed proto-planet would be losing mass via a Parker wind that is  {\bd catalyzed} by the stellar continuum radiation. {\bd This represents an intermediate stage of mass-loss, occurring soon after the disc has dispersed, but before { the EUV/X-ray driven} photoevaporation becomes relevant.}  The surface mass-loss induces a mass movement within the envelope that advects internal heat outward. As a result, the planet atmosphere rapidly cools down and contracts, until it has reached a radius of order $0.1$ Bondi radius, at which time the mass-loss effectively shuts down. Within a million years after the disk disperses, we find a planet that has only about { ten}  percent of its original envelope, and a Kelvin-Helmholtz time that is much longer than its actual age. We suggest that this ``boil-off'' process may be {\bd partially} responsible for the lack of planets above a radius of $2.5 R_\oplus$ in the {\it Kepler} data, {\bd provided planet formation results in initial envelope masses of tens of percent.}
\end{abstract}





\section{Introduction}
The {\it Kepler} observatory has revealed the presence of a large number of close-in ($\lesssim 1$~AU), small ($\lesssim 4$~R$_\oplus$) \citep{Borucki2011,Batalha2013a,Mullally2015} and low-mass ($\lesssim 20 M_\oplus$) exoplanets {\bdd \citep{Marcy2014,Dumusque2014,Dressing2015}}. In fact this type of planet is so frequent most stars are thought to contain one \citep[e.g.][]{Fressin2013,Petigura2013,Silburt2015}. We nick-name these planets ``{\it Kepler} planets''. They may represent the dominant mode of planet formation, allowing us to calibrate our planet formation models and finally understand this illusive problem.

Unlike the terrestrial planets in our own solar system, the inferred masses and radii of  the {\it Kepler} planets indicates that a large fraction contain H/He rich envelopes \citep{Wu2013,Hadden2014,Weiss2014a,Rogers2014a}. Even among ones that appear to be naked cores, it has been argued that hydrodynamical escape may have occurred and removed their primordial envelopes  \citep{Owen2013b,Lopez2013}. This indicates that, in most cases, planet assembly must have finished by the time gas-rich protoplanetary disks disperse after a few Myr.

The size distribution of {\it Kepler} planets exhibits a puzzling feature: there appears to be a concentration of planets with sizes $\sim 2.5 R_\oplus$, with a steep fall-off in number at larger sizes, and a  possible reduction in number toward smaller sizes {\bdd \citep{Petigura2013,dfm2014,Silburt2015}, although the quantitative details are still uncertain}. For the mass range of interest, a size of $2.5 R_\oplus$ corresponds to an envelope mass fraction of $\sim 1\%$ {{\bd \citep{Wu2013,Wolfgang2015a}}}. Why does nature favour such an envelope mass, when the environments (e.g., gas density, temperature) for planet formation may be highly diverse and time varying?

There are currently two schools of thought for where these planets formed:  either  they are formed ``in-situ'',  close to their current small orbital separations \citep[e.g.][]{Hansen2012,Chiang2013}, or they formed at larger separations and  are then migrated to  current positions \citep[e.g.][]{Ida2005,Ida2010,Bodenheimer2014c,Raymond2014}. Even in the latter scenario, the final planet assembly and the gas accretion may have occurred near the host stars. {\it Kepler} planets that are in high multiple systems appear to be so closely packed \citep[e.g., the Kepler-11 system,][]{Lissauer2011,Mahajan2014b}, they would have been skirting dynamical instabilities in the final assembly had they not been subject to substantial eccentricity dissipation \citep{Pu2015}. Their current low values of eccentricities \citep{Wu2013,Hadden2014} also attest to this dissipation. Gas is the most obvious source of this damping.

Could the gas accretion process, at a distance of $\sim 0.1$ AU from the host star, naturally give rise to the above-mentioned $1\%$ envelope mass? The answer is no, at least not according to current theories. In the theory of core-accretion \citep[e.g.][]{Pollack1996}, a planetary embryo embedded in a gas disk will quickly acquire a hydrostatic envelope extending out to its Bondi sphere\footnote{In the inner disk{,}  the Bondi radius is smaller than the Hill Sphere for the core masses we are  {\bd concerned} about \citep[e.g.][]{Rafikov2006}.}, with a radius\footnote{
It is yet unclear exactly where the proto-planet's envelope ends and where the disk begins.  Background shear in the disk can affect the transition radius \citep{Lissauer2009}, as can 3-D flow dynamics \citep{Ormel2015,Fung2015}. We ignore these complications here.}
\begin{equation} R_B = {{G M_p}\over{2 c_s^2}} \, ,
\label{eq:rb}
\end{equation}
where $M_p$ is the proto-planet's mass and $c_s$ is the isothermal sound speed of the surrounding gas. The atmosphere, confined by an external disk pressure, has a mass that is roughly the background gas density times the volume of the Bondi sphere \citep{Rafikov2006}.  If this is the final atmosphere that we observe, its mass should depend on gas density, temperature and embryo mass. The observed clustering around $1\%$ mass fraction therefore comes as a surprise.

Multiple works have suggested that the above mass estimates may not be the final envelope mass. \citet{Ikoma2012} argued that, as the gas disk gradually dissipates away, lifting of the pressure confinement will slowly erode away the envelope. In contrast, \citet{Lee2014} asserted that due to a lack of planetesimal bombardment and associated heating for embryos in the inner disk, the above mentioned hydrostatic envelope will cool and contract, allowing the embryo to accrete ever more gas. In fact, they found that even a $5 M_\oplus$ planet may accrete so much gas as to undergo unstable run-away growth \citep{Mizuno1978}. We return to comment on these works later in the discussion.  Instead, in this work, we focus on a different aspect of the problem, namely, the case where the nascent disk disappears quickly {(as indicated observationally, e.g. \citealt{Koepferl2013})}, leaving the newly formed planet exposed to stellar irradiation and a vacuum boundary condition. We argue that this brings about vigorous mass-loss and rapid cooling, strongly impacting the final envelope mass. This process, happening in the final stage of low-mass planet formation, may potentially help explain the $1\%$ convergence one observes in {\it Kepler planets}. 

{\bd This new stage of mass-loss is distinctively different from the process discribed in \citet{Ikoma2012} and from the mass-loss induced by EUV/X-ray photoevaporation \citep{Lopez2013,Owen2013b}. As our calculations show (\S 3), it occurs because the planet is highly inflated and because it is strongly illuminated by stellar photons. Both these conditions may be satisfied for low-mass planets that newly emerge out of the protoplanetary disk, due to their long thermal timescales (\S 2).  The mass-loss takes the form of a Parker wind \citep{Parker1958} that is heated by continuum stellar radiation (as opposed to only high energy photons). But the ultimate energy source is the internal heat of the planet. As a result of envelope removal, the planet cools and contracts quickly (\S 4). We consider the implications of this work in \S 4 and conclude in \S 5. }

{\bd In this work we concentrate purely on the case of a rocky core surrounded by a H/He envelope, i.e. the structure most likely to arise from planets forming close to their star in a gaseous disc, although we note many other structures are still consistent with the observed planet population \citep[e.g.][]{Rogers2010}.} 

\section{Initial Conditions and Presence of Mass-loss}\label{sec:setup}

Planets can lose their atmospheres { over time}. This process is more extreme when the planets are strongly irradiated and highly inflated. We argue here that this may be just the {\bd thermal} state the {\it Kepler} planets {\bd found themselves in,} when their parent disks dissipated. 
\x{{\bd We are specifically interested in the case after the gas disk has dispersed and the planet effectively sees a vacuum boundary condition and is irradiated by the central star, but before photoevaporation had time to have an effect.}}
{\bd The process we are interested in occurs within $10^4-10^5$ yrs after the disk dispersal, before the EUV/X-ray photoevaporation has had much of an effect. The latter occurs on a time-scale of $10^8$ yrs \citep{Owen2013b}.}

\subsection{ A wind is driven}

The black-body temperature at {\bd a} distance $a$ from a star with surface temperature  {\bd $T_*$} and radius $R_*$ is
\begin{equation}
T_{\rm eq}=886\,{\rm K}\left(\frac{T_*}{5800\,{\rm K}}\right)\left(\frac{R_*}{1~{\rm R}_\odot}\right)^{1/2}\left(\frac{a}{0.1\,{\rm AU}}\right)^{-1/2}\, .
\label{eq:Teq}
\end{equation}
This is also the photospheric temperature of a planet {\bd when its} internal luminosity is subordinate to stellar irradiation {\bd and when it is exposed to direct sun-light}.  {\bd In this paper, we assume} that the gas above the photosphere, {\bd in particular, the extended atmosphere that is the outflow we are investigating,} {\bd will} also {\bd be} heated to roughly this temperature. {\bd This could happen due to the} absorption of continuum photons by the dust grains embedded in the outflow, {\bd when they are present,} or by absorption of ionizing photons by the gas molecules when dust is absent. {\bd The planet distances of concern lie beyond the dust sublimation radius. So we expect dust formation to be efficient in the outflow. In this work, we do not attempt to model the temperature profile in the extended atmosphere, a short-coming we hope to remedy in future works. We do remark that the
condition of isothermal wind sets an upper bound to the mass-loss rate: the energy advected by the wind should not exceed stellar energy deposit in the wind, or else the isothermal condition will break.}  

An isothermal atmosphere has an outwardly increasing scale height and can only be in hydrostatic equilibrium  if the ambient pressure is sufficiently strong to confine it. This confining pressure was provided by the proto-planetary disk while it was around, but once it dissipated, the only candidate is the stellar wind, which, even at the phenomenal mass-loss rate of young stars \citep[e.g.][]{Wood2002}, falls  far too short to be of use. So the isothermal atmosphere is not hydrostatic but instead {harbours} a
so-called ``Parker wind'' \citep{Parker1958}, with a sonic radius at:
\begin{equation}
R_s=\frac{GM_p}{2c_s^2}=R_B\, .
\end{equation}
Note the fact that the sonic radius and the Bondi radius are identical is not accidental, but represents the symmetry between outflow and accretion. 

{\bd Energetically speaking, the stellar continuum radiation, which maintains the isothermality of the wind, is the reason for the mass-loss. However, as we will discover in later parts, the mass-loss brings about a rapid gravitational contraction of the planet. The binding energy released during this contraction fuels the mass-loss. There are thus two energy sources for the mass-loss, with the bottom (planet interior) pushing material up, and the top (stellar heating) pulls the material further away from the planet.}

\begin{figure}
\centering
\includegraphics[width=1.1\columnwidth]{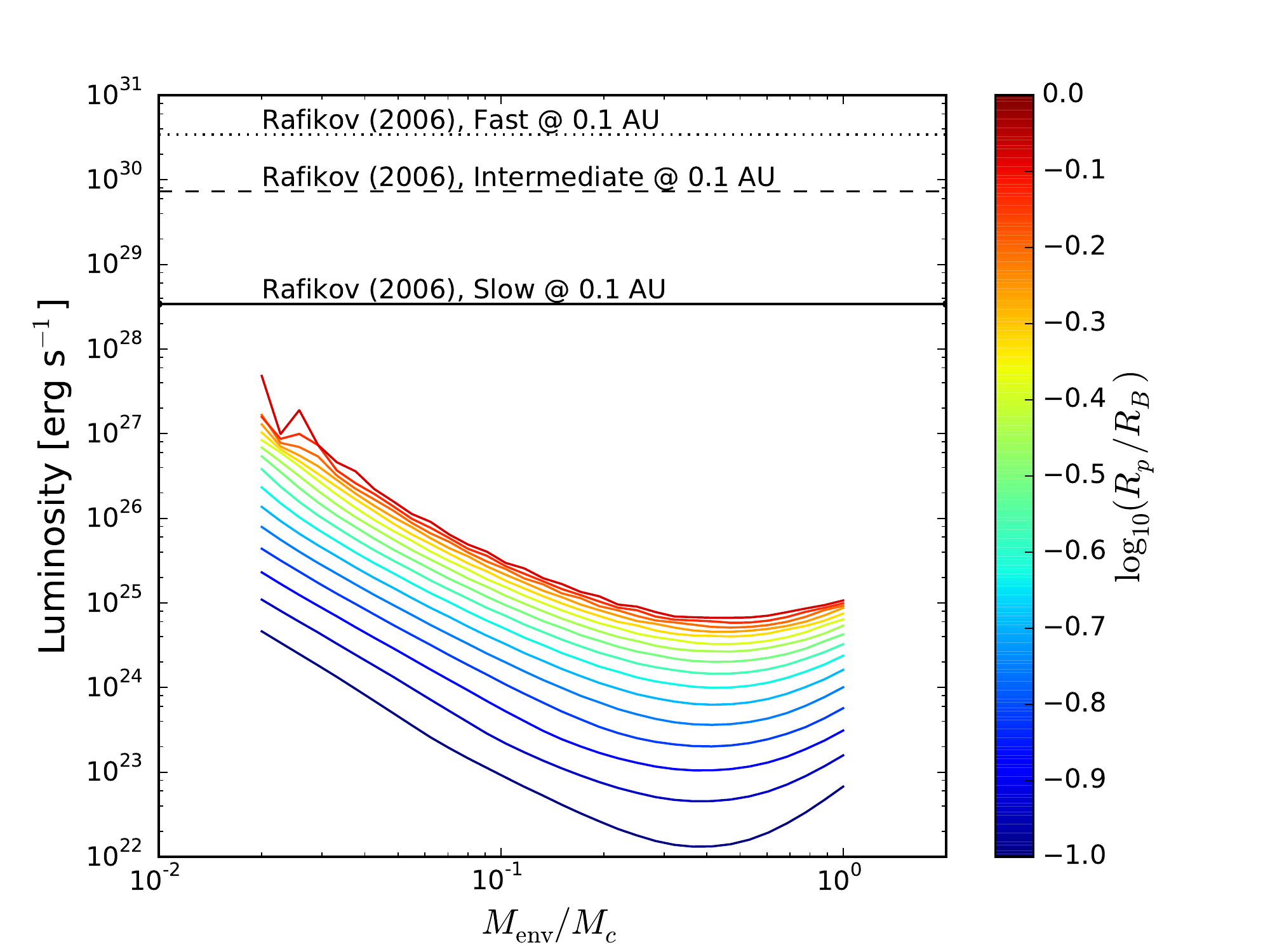} 
\caption{The internal luminosity of a planet with a $5 M_\oplus$ rock core \x{with} { and} a surface temperature of $900\K$ (roughly { the} blackbody temperature at $0.1$AU) { as a function of the envelope mass fraction}.\x{, and variable envelope mass for fixed radius (in units of the Bondi-radius).} The colour of the lines indicates the radius of the planet in units of the Bondi-radius \x{ranging from roughly 0.1 to 1.0}.
The three horizontal lines label the expected accretion luminosities from solids in a MMSN disk \x{taken from} \citep{Rafikov2006}. { At these values, solid accretion} easily overwhelm the cooling contraction of the proto-planets.\x{, if these solids are present at the MMSN level.}}\label{fig:figure1}
\end{figure}

\subsection{Race between Cooling and Dispersal}

{\bd The importance of our new mechanism depends critically on the size of the planet when it was first exposed to direct stellar irradiation.}
A highly inflated planet, with the photosphere at its Bondi radius, will experience a remarkable mass-loss rate of:
  \begin{eqnarray}
  \dot{M}_p&\sim &4\pi R_B^2\rho_{\rm surf}c_s=4\pi R_B^2\frac{P_{\rm surf}}{c_s} \nonumber \\
  &\approx& 1\times10^{-2} \,{\rm M_\oplus\,yr^{-1}}\,\left(\frac{M_p}{10\,{\rm M}_\oplus}\right)\left(\frac{c_s}{2.0~{\rm km~s}^{-1}}\right)^{-1}\nonumber \\ &\times& \left(\frac{\kappa}{0.1\,{\rm cm^2 g}^{-1}}\right)^{-1}\, , \label{eq:mdotrb} \end{eqnarray} 
where we have taken the photosphere pressure as $P_{\rm surf} \approx g/\kappa$ with $g$ being the local gravitational acceleration, $\kappa$ the opacity {\bd and the chosen sound speed corresponds to a separation of $\sim 0.1$~AU for molecular hydrogen}. Clearly,  such a mass-loss rate is not sustainable and it will bring dramatic changes to the planetary atmosphere, an issue we develop later in this paper. For the moment, we only consider the on-set of mass-loss. The mass-loss rate drops off exponentially with smaller planet sizes, as the sonic point density depends exponentially on planet sizes \citep{Cranmer2004c}. So the importance of our new process depends on the size of the planet when the disk is removed.

In turn, how inflated a newly emerged planet is, depends on the race between the cooling/contraction of its envelope and the {\bd external} process of disk dispersal.  If the cooling time-scale is much shorter than the disk dispersal time-scale, then the planet\x{s} will contract significantly, rendering post-partum mass-loss insignificant.

\begin{figure*}
\centering
\includegraphics[width=\textwidth]{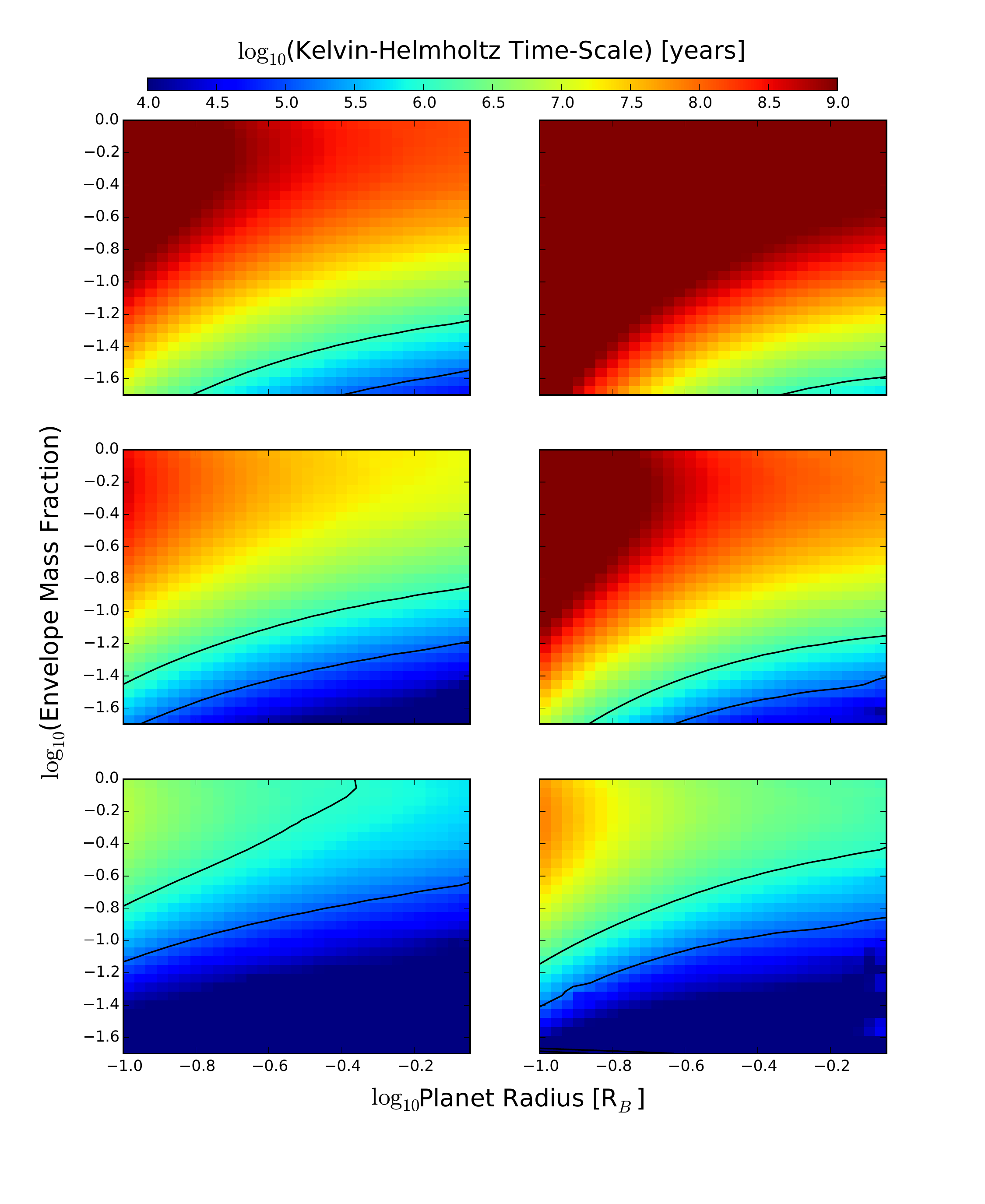}
\caption{The Kelvin-Helmholtz time-scale as a function of planet radius in units of the Bondi radius and envelope mass fraction. The solid contours show Kelvin-Helmholtz time-scales of $10^6$ and 10$^5$ years. The left panels show planets with an equilibrium temperature of 500~K and the right panels show planets with an equilibrium temperature of 900~K. From top to bottom the panels show planets with core masses of 3, 5 and 10 M$_\oplus$. }\label{fig:tkh}
\end{figure*}

The relevant cooling time-scale of a gravitational bound object with no internal energy sources (e.g. nuclear burning) is given by the Kelvin-Helmholtz time-scale ($t_{\rm KH}$), which for a planet with envelope mass ($M_{\rm env}$) less  than the core mass ($M_c$)  is:
\begin{equation}
t_{\rm KH}\approx \frac{GM_cM_{\rm env}}{R_pL}
\end{equation}
where $L$ is the planet's cooling luminosity. For an embedded planet, one can assume that any cooling flux that reaches the Bondi radius is instantly removed by the external shear flow. This is similar to a planet exposed to vacuum. So the actual cooling luminosity only depends on the internal structure of the planet, not  on its environment.

 The near-surface temperature structure determines the cooling luminosity. In the deep atmosphere, heat is mostly advected outward by convective eddies; near the photosphere, this could be accomplished by radiation diffusion. Thanks to stellar heating, the surface temperature of the planet is equilibrated to that of the local blackbody (eq. \ref{eq:Teq}) if it is exposed, or to the mid-plane disk temperature if it is embedded.   This surface searing tends to flatten out the surface temperature gradient to become more isothermal, reducing the heat escape.{ This situation  is similar to that for an irradiated hot jupiter \citep{Guillot1996,Burrows2000,Arras2006}.
 
The thickness of the radiative zone depends on envelope mass, among other things.  { The planet atmospheres are fully adiabatic if their internal entropy is as high as the photospheric value-- these} are the hydrostatic primordial atmospheres as calculated by \citet{Pollack1996,Rafikov2006}. Because of their large scale height, they contain the least { possible} amount of mass.  To load more mass into these atmospheres, the interiors have to cool to a lower entropy than \x{that in the disk} { the surface value}. { The lowering of the internal temperature, allied with the surface searing, then entails an ever thickening isothermal blanket.} \x{ The convective interior then needs to be covered by a largely isothermal blanket, before contact with the disk. }  This isothermal layer is effective at blocking the internal heat flow. 

As a result, for planets with the same core mass, size and surface temperature, the cooling luminosity is the largest in ones that have fully convective atmospheres { (and the lowest envelope masses)}, while it decreases rapidly in ones with an isothermal upper layer { and {\bd accordingly} larger envelope masses,}   as is seen in Fig. \ref{fig:figure1}, obtained for planets with a $5 M_\oplus$ rock core with a surface temperature of $900\K$.\footnote{The photospheric pressure { in these models} may differ from that at the disk mid-plane, making these models imperfect for embedded planets. However, luminosity from such a model is still well characterized as it is largely determined at the radiative-convective boundary \citep{Arras2006,Wu2013}.} The Kelvin-Helmholtz cooling times, for a wide range of models, are presented in Fig. \ref{fig:tkh}, where we vary planet photosphere radius (in unit of Bondi radius), envelope mass fraction, surface temperature and core mass.  These values are calculated using the {\sc mesa} stellar and planetary evolution code \citep{Paxton2011,Paxton2013a}, using the setup described in \S4.  All else being equal, one observes that models with { a} higher envelope mass cool\x{s} much slower. In addition, higher stellar insolation slows down the cooling contraction. Lastly, cooling is reduced in models { that} are less extended in size. All of these trends can be explained by the size of the isothermal blanket.

We now turn to discuss the time-scale of disk dispersal. There are two time-scales of relevance. The first is the time-scale a fully grown planet remains embedded, or the remaining disk lifetime after the planet is formed. Since we do not know when planets are formed, we adopt an upper limit to this value, i.e., the full disk lifetime, $\sim 3-10$ Myrs \citep[e.g.][]{Haisch2001,Hernandez2007,Mamajek2009}.
The second is the time-scale the disk density drops from its full value to nearly zero. As has become clear { by} now, the two { time-scales} are not necessarily the same.  Disk dispersal -- particularly in the inner disk -- is known observationally \citep[e.g.,][]{Kenyon1995,Ercolano2011e,Koepferl2013} to be at least an order of magnitude quicker than the disk's lifetime, { the} so called ``two time-scale'' disk evolution. These observations have lead to the development of the photoevaporative-switch dispersal model \citep{Clarke2001,Alexander2006,Gorti2009,Owen2010,Owen2011e} where photoevaporation of the outer disk carves a gap around 1~AU (for a solar mass star), sabotaging the gas supply to the inner region, causing the latter to rapidly drain onto the central star on its short {\it local} viscous time-scale, which is typically $10^5$~years. Alternatively, if the planet is massive enough to open { a} gap, the { local disk clearing} time-scale can be as short as a few orbital periods \citep[e.g.][]{Crida2006}.

Having established the time-scales for cooling contraction and disk \x{time-scales} { dispersal}, we can now assess the size of newly emerged planets. {  Fig. \ref{fig:tkh} shows } that a $5 M_\oplus$ planet can remain at a size $R_p \sim R_B$ well after disk removal, if its envelope mass fraction is more than $\sim 10\%$. This threshold reduces to $\sim 5\%$ for a $3M_\oplus$ planet, while a $10 M_\oplus$ planet can not remain inflated at this size for any relevant envelope mass (this is not unexpected as 10~M$_\oplus$ cores are those typically required for giant planet formation e.g. \citealt{Rafikov2006,Piso2014} in the outer disk).  The last observation is consistent with the finding of \citet{Lee2014}, where they argue that the envelopes of these planets cool and contract so rapidly, they should undergo runaway gas accretion even while embedded.

{ Further, to this discussion, one may argue that an embedded planetary core continuously accretes more envelope until the envelope cooling time becomes comparable to the 
disk lifetime. Adopt a value of 1 Myrs, this corresponds to an envelope mass of $\sim 
7\%$ for a $5M_\oplus$ planet at 0.1 AU.}

 In these discussions, we have overlooked an important detail. We have assumed that the only planet luminosity comes from its cooling contraction. In reality, there is the possibility that the planet can accrete planetesimals (solids) from the surrounding disk. If this is substantial, it could maintain the high entropy of the gas envelope and keep the planet inflated, well past its Kelvin-Helmholtz time-scale. As is shown in Fig. \ref{fig:figure1},  reasonable estimates for the accretion luminosity \citep[e.g.][]{Rafikov2006} easily overwhelm the cooling luminosity {\bd and may keep the planets inflated} 
longer than that estimated in Fig. \ref{fig:tkh}.

 In summary, we argue that newly emerged low-mass planets ($M < 10 M_\oplus$), {  fully assembled within the inner gaseous disk and have} \x{ with}  non-negligible envelope masses ($\sim5 - 30\%$), will remain inflated ($R_p \sim R_B$) when the disk disappears {\bd in a rapid timescale of $\sim 10^5$ yrs}, and are thus susceptible to strong  winds. 
s{\bd This stage of mass-loss occurs after the type of mass-loss studied by \citet[e.g.][]{Ikoma2012}, and before the 
EUV/X-ray photoevaporation \citep[e.g.][]{Owen2013b} that lasts for 100 Myrs.}
We dub this process the ``boil-off'' and proceed to study it schematically in \S3 and numerically in \S4.
\x{ noting that such a scenario is inevitable if the {\it Kepler} planets finished their assembly in the inner gaseous disk}. 

\section{The ``Boil-Off'' and its consequences}\label{sec:mdot}

{ We now consider the rate of mass-loss and the cooling contraction for a low-mass planet, irradiated by a strong stellar flux after it emerges from the nascent gas disk.}

\x{ that has left the surrounding disk and is irradiated with a radiation field giving a photosperic temperature equal to the equilibrium temperature and that the internal luminosity of the planet is much smaller than the stellar flux absorbed. Thus, the planetary structure can be approximated as convective from the core radius until the envelope temperature drops to $T_{\rm eq}$, then the temperature profile will then be isothermal at $T_{\rm eq}$.}

\subsection{Mass-loss rates}
\label{subsec:massloss}

A highly inflated young planet, exposed to the stellar radiation, is vulnerable to mass-loss. {\bd If the wind remains isothermal, {\bd as we have assumed,} then the appropriate velocity and density profile is given by  the Parker wind model \citep{Parker1958}.} 
 
Eq. \refnew{eq:mdotrb}, generalized for any planet size $R_p$ with $R_p < R_B$, becomes: \begin{eqnarray} \dot{M}&=&4\pi R_p^2 \rho_{\rm surf} u_{\rm surf}
  = 4\pi R_p^2\mathcal{M}_p \left(\frac{P_{\rm surf}}{c_s}\right)\nonumber \\
  &=& \frac{4\pi G M_p}{\kappa c_s}\mathcal{M}_p \, , \label{eq:mdot_1} \end{eqnarray} where $\mathcal{M}_p$ is the Mach number of the flow at the planet's photosphere and is only a function of the ratio $R_p/R_B$, given by: \begin{eqnarray}
  \mathcal{M}_p&=& \sqrt{-W_0\left[-f(R_p/R_B)\right]}\label{eqn:mach_parker}\\
  &\approx& \left(\frac{R_p}{R_B}\right)^{-2}\exp\left(-\frac{2R_B}{R_p}\right) \mbox{ when $ R_p \ll R_B$}\, , \label{eqn:expand} \end{eqnarray} where $W_0$ is a real branch of the Lambert function \citep{Cranmer2004c} and \begin{equation} f(x)=x^{-4}\exp\left(3-\frac{4}{x}\right)\, .  \end{equation} {\bd We verify that the energy required to heat up the wind, given the mass-loss rate (eq. [\ref{eq:mdot_1}]), is subordinate to the stellar irradiation{\bd for an equilibrium temperature of 1000~K}, as is shown in Figure~\ref{fig:L_compare} for a 10 and 5 earth mass planet.  So energetically, the wind can be heated to isothermal by the star. {\bdd This energy argument, however, breaks down at an equilibrium temperature of $\sim 250$~K for a 5 earth mass planet and $\sim 200$~K for a 10 earth mass planet. Therefore, at large separations,  $\gtrsim$ 1~AU around a Sun-like star, the flow cannot remain isothermal, leading to lower mass-loss rates. }

  \begin{figure}
    \centering
    \includegraphics[width=\columnwidth]{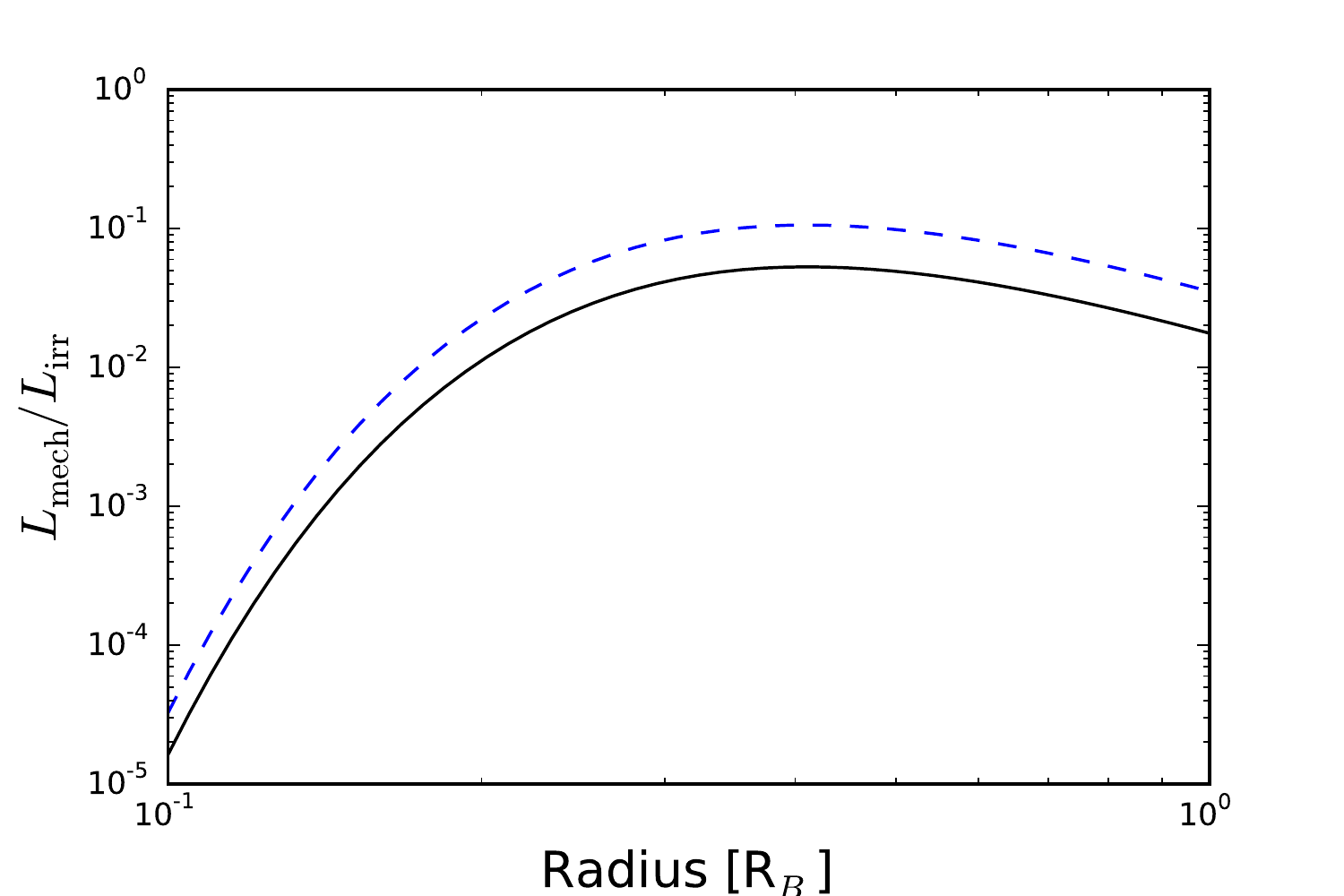}
    \caption{Figure showing the ratio of mechanical luminosity in a Parker wind, compared to the recived energy from radiation as a function of planet radius. The solid line shows a 10~M$_\oplus$ planet and the dashed line shows a 5~M$_\oplus$ planet both with an equilibrium temperature of 1000~K. The mechanical luminosity is considerably smaller than the energy  from irradiation. }\label{fig:L_compare}
  \end{figure} 

  Eq. \refnew{eq:mdot_1} can be used to calculate a mass-loss time-scale
  \x{ ($t_{\rm ML}=M_{\rm env}/\dot{M}$)} as:
  \begin{eqnarray}
    t_{\rm ML}&{ \equiv} & {{{M_{\rm env}}\over {\dot{M}}}} =\frac{\kappa c_s}{4\pi G}X_{\rm env}\mathcal{M}_p^{-1}\nonumber \\
    &\approx& 10^3\,\mbox{yrs}\;X_{\rm env}\mathcal{M}_p^{-1}
    { \left({{T_{\rm eq}}\over{886\K}}\right)^{1/2}}
    \left(\frac{\kappa}{0.1\,{\rm cm^2 g}^{-1}}\right)\, ,
  \end{eqnarray}
  where $X_{\rm env}$ is the envelope mass fraction. In Fig. \ref{fig:tmdot} we illustrate how the Mach number at photosphere, as well as the mass-loss time-scale, depend on the ratio of $R_p/R_B$, by solving the exact Lambert function. But here, we provide a schematic argument for the core result.
  Using the expression for the Mach number when $R_P \ll R_B$ (eq. \ref{eqn:expand}), we can recast the above equation as:
  \begin{eqnarray} &&\left(\frac{R_p}{R_B}\right)^2\exp\left(\frac{2R_B}{R_p}\right)\approx X_{\rm env}^{-1}\left(\frac{t_{\rm ML}}{10^3{\rm \, years}}\right)\nonumber \\&&\times 
    { \left({{T_{\rm eq}}\over{886\K}}\right)^{-1/2}}
    \left(\frac{\kappa}{0.1\,{\rm cm^2 g}^{-1}}\right)^{-1}\, .
    \label{eqn:sol} \end{eqnarray}
  So at small $R_p/R_B$,  the \x{RHS}  mass-loss is  exponentially sensitive to the value of $R_p/R_B$. Alternatively, one can say that  the final radius where mass-loss effectively stalls is only logarithmically sensitive to the exact model parameters, and is always of order $R_p/R_B \sim 0.1$. As can be observed in Fig. \ref{fig:tmdot}, this is indeed the point where the mass-loss time-scale sky-rockets to exceptionally long time-scales. 

  \begin{figure}
    \centering
    \includegraphics[width=\columnwidth]{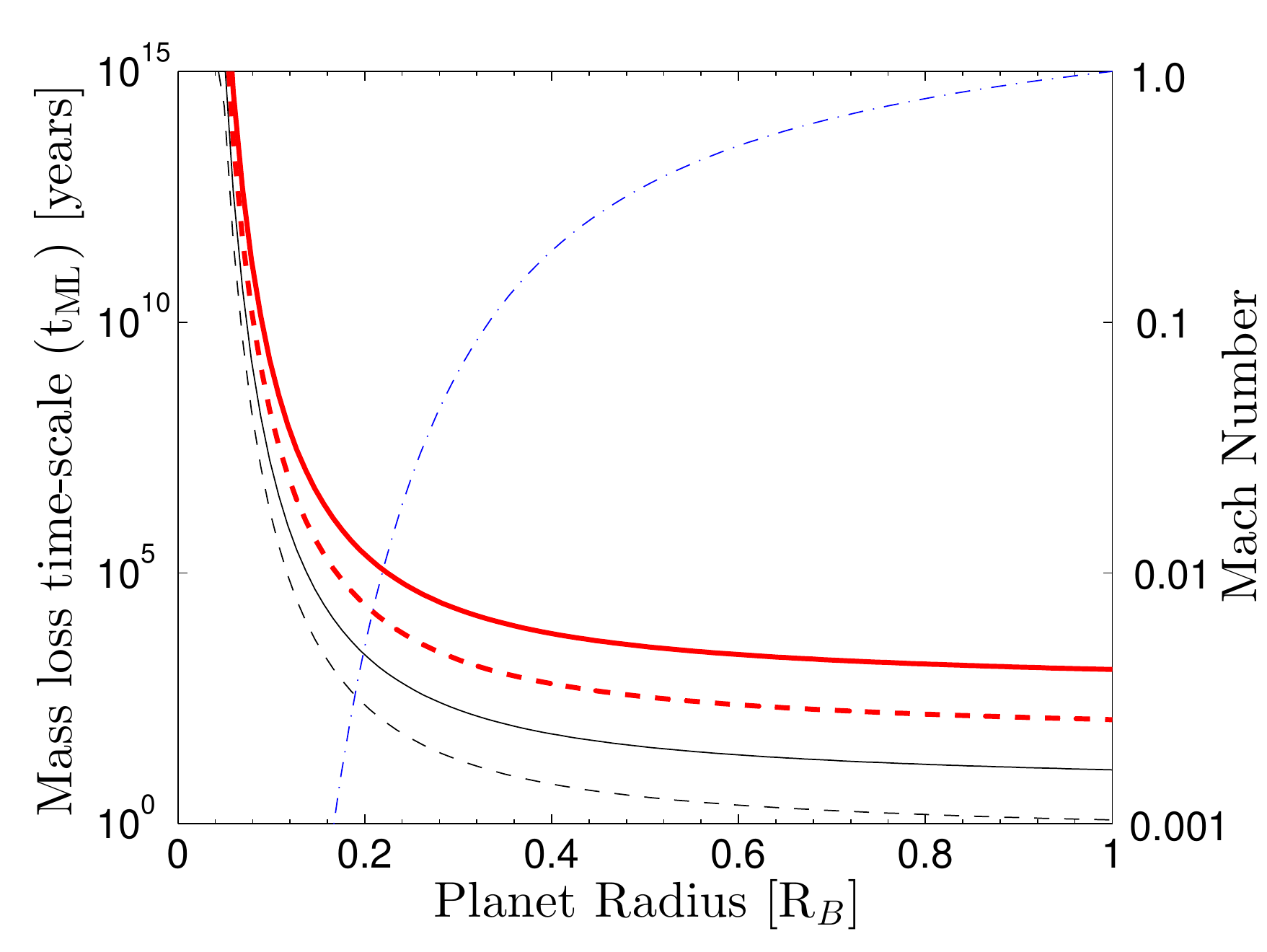}
    \caption{Mass loss time-scale ({\bd left Y-axis}) and launch Mach number ({\bd right Y-axis}) as a function of planet radius normalised in terms of the Bondi radius. Thick red lines show the mass-loss time-scale for models with $\kappa=0.01$ cm$^{2}$ g$^{-1}$ whereas thin black lines show the mass-loss time-scale for models with $\kappa=1$ cm$^{2}$ g$^{-1}$; dashed lines indicate models with $a=0.1$~AU and solid lines with $a=1.0$~AU. The thin blue dotted-dashed line shows the launch Mach number. {\bd We note the two Y-axes are not linked directly.} }\label{fig:tmdot}
  \end{figure}
  So, once the confining pressure of the disk is  lifted, the planet's atmosphere - although in roughly dynamical equilibrium - will undergo rapid mass-loss. Such a mass-loss, if happening at a pace faster than the intrinsic cooling contraction of the planet, as is indeed the case in Fig. \ref{fig:tmdot}, will dominate the thermal evolution of the planetary atmosphere. \x{As we show below, this ``boil-off'' carries away so much of the internal energy, the planet cools and contracts catastrophically. This shrinkage, in turn, effectively shuts off the mass-loss when $R_p$ has dropped down to $\sim 0.1 R_B$. The ``boil-off'' phase is then terminated. }

  \subsection{Consequences for thermal evolution}

  Just like evaporation from the skin cools us in the summer heat, the rapid mass-loss from young planets will lead to dramatic cooling and contraction.

  In a planet without mass-loss, its internal heat is transported outward by a combination of convection and radiative diffusion. The heat leakage, and hence the rate of cooling contraction, is limited by the bottle-neck in this transport process, i.e., the outer radiative zone. This is the reason behind the results shown in Fig. \ref{fig:tkh}.{ Mass-loss alters this picture.} The outflow advects thermal energy at a rate:
  \begin{eqnarray}
    L_{\rm adv}&\approx&\frac{\gamma}{\gamma -1}\dot{M}c_s^2\sim 10^{26} {\rm \,erg\,s^{-1}}\,\, \left(\frac{\dot{M}}{1\times10^{-5} \,{\rm M_\oplus\,yr^{-1}}}\right) \nonumber \\ &\times &\left(\frac{c_s}{3\times10^{5}\,{\rm cm\,s^{-1}}}\right)^2\, ,
  \end{eqnarray}
  where $\gamma$ is the ratio of specific heats. This flux easily overwhelms  the cooling luminosities of low-mass planets (Fig. \ref{fig:figure1}).
  As a result, the hot interior of the planet can shed its energy at a rate much higher than that permitted by the radiative bottle-neck. The planet cools down in a hurry.

  This cooling reduces the thermal support for the atmosphere and it contracts. The contraction reaches $R_p \sim 0.1 R_B$ within a few $t_{\rm ML}$. This effectively shuts off the mass-loss and cooling resumes the original pace in which the radiative bottle-neck is important {\bd \citep{Owen2013b,Lopez2013}}. Unlike the case of no mass-loss, where $t_{\rm KH}$ is always of order the age of the planet, we now have a planet which is young in age, but { ``looks'' old. Specifically, it has a Kelvin-Helmholtz time-scale that is considerably larger than its age.}

  \subsection{Final envelope mass}\label{sec:final_mass}

  \x{While the final planetary radius at the end of {\y the ``boil-off''} can be easily estimated, the final mass of the planetary atmosphere is harder to estimate and will depend on the detailed thermal evolution \x{ during this phase}. We proceed to consider this \x{below} {\y here}.}

  Knowing the initial and final radius of the planet envelope, we can estimate { the amount of mass-loss} using  an energy argument. 

  For the initial state, the binding energy of the envelope is:
  \begin{equation}
    U_i=-A_i\frac{GM_{c}M_{\rm env}^i}{\alpha R_B}\, ,
  \end{equation}
  where $A_i$ is an order unity constant  and is determined by the central concentration in the envelope. $\alpha$ is a parameter ($\le1$) that represents the initial radius of the planetary atmosphere in terms of the Bondi radius.  After the episode of mass-loss, the envelope has shrunk to a radius of $0.1 R_B$ and the binding energy of the envelope is now:
  \begin{equation}
    U_e=-10A_e\frac{GM_cM_{\rm env}^f}{R_B}\, .
  \end{equation}
  where $A_e$ is also an order unity number that describes the central concentration of the new state. As the mass-loss rate is exponentially sensitive to planet radius, we expect that most of the loss occurs when the planet was at its initial size, and this requires a binding energy release of:
  \begin{equation}
    U_{\rm lost}  \approx \frac{GM_c\left(M_{\rm env}^i-M_{\rm env}^f\right)}{\alpha R_B}\, .
  \end{equation}
  Assuming that the potential energy difference between the initial and the final state is all used to drive the Parker wind, or, $U_i - U_e = U_{\rm lost}$, we can solve for the final envelope mass as:
  \begin{equation}
    M_{\rm env}^f=\left(\frac{A_i+1}{10\alpha A_e+1}\right)M_{\rm env}^i \approx 0.2 \alpha^{-1} M_{\rm env}^i \label{eqn:menv_f1},
    \label{eq:mfinal}
  \end{equation}
  where we have taken $A_i \sim A_e \sim 1$.

  In this discussion, we ignore two other energy terms. One is the energy gain from the stellar insolation, which can be orders of magnitude larger than the planet's internal luminosity; and one is the energy loss from convective/radiative transport. The latter, as we have argued, is subordinate to the advective energy flux due to mass loss and can be safely ignored (also see numerical evidences in \S \ref{sec:numerical}). The first assumption, however, takes some considerations.

  First, stellar heating is essential to drive the Parker wind. It continuously heats the gas above the photosphere, allowing them to reach escape velocity. Without it, the atmosphere will happily remain at a hydrostatic equilibrium. However, stellar heating is only a minor energy source  for mass-loss, not the driving engine.  Stellar heating provides the final push to take a gas parcel from the photosphere to infinity, and the
  pressure loss caused by the outflow allows gas to rise up from below. The increase in {its} binding energy is provided by the gravitational contraction of the underlying gas. {\bdd If we include the energy gain due to stellar irradiation into our simple analysis, $U_{\rm irr}=\epsilon(R_B/a)^2L_*t_{\rm boil}$, where $L_*$ is the stellar luminosity, $t_{\rm boil}$ is the length of the boil-off phase and $\epsilon$ a small pre-factor $\ll 1$, we find,}
  \begin{equation}
    M_{\rm env}^f\approx 0.2 \alpha^{-1} M_{\rm env}^i -\epsilon\left(\frac{R_B}{a}\right)^2\frac{L_*t_{\rm boil}R_B}{GM_c}\, .
  \end{equation}
  {\bdd This shows that stellar irradiation can increase the mass-loss slightly, with the strongest effects for small separations and low core masses. }

{\bd In numerical simulations, we frequently find that the final envelope mass is some $10\%$ or less of the initial mass (Fig. \ref{fig:5Me_evolve}), {\bdd indicating that stellar irradiation is a sub-dominant energy source.}}

\section{Numerical calculations}
\label{sec:numerical}

We simulate the thermal and dynamical evolution of a spherically symmetric { planetary envelope}, subject to the boundary conditions of stellar irradiation and mass-loss. The gas is optically thick to radiation ($\tau \geq 2/3$) so that the radiation field and the fluid are in local thermal equilibrium. Additionally, we assume that the radiation transport  takes place in the static diffusion limit, i.e., $v/c\ll \lambda_p/\ell$ (where $\lambda_p$ is the mean free path of the photons and $\ell$ is the scale length of the atmosphere), a condition easily satisfied in our problem}. 

\subsection{Numerical method}
\label{subsec:code}
 We use the {\sc mesa} code \citep{Paxton2011,Paxton2013a} to solve this radiation-hydrodynamics problem.  This code traces the evolution of any mass shell inside the planet, as is dictated by the equations of mass conservation, momentum conservation and energy conservation, subject to suitable boundary conditions. 
This is valid so long as the simulated domain contains no shocks and sonic points.
 In other words, we can not follow the Parker wind through its escape process. Instead, we have to independently specify a mass-loss rate at the outer boundary of the planet.
 Since {\sc mesa} is essentially a Lagrangian hydrodynamics code { (namely, following the mass element), this is not straightforward to implement and we discuss this below.

\subsubsection{Implementation of mass-loss}\label{sec:mdot_implement}

 To specify the mass-loss rate in {\sc mesa}, one removes a certain amount of mass from the top layers (we take the surface to be the photosphere with $\tau = 2/3$ {to the outgoing radiation})  at the beginning of  each time-step  and then lets the system readjust, as is described in detail in the {\sc mesa} code paper \citep{Paxton2011}.   

In our implementation, we assume that the flow above the photosphere follows the isothermal Parker wind solution with a temperature equal to the equilibrium temperature $T_{\rm eq}$ { (Eq. \ref{eq:Teq})}and a mass-loss rate as in eqs. \refnew{eq:mdot_1}-\refnew{eqn:mach_parker}.
\x{\begin{equation}
\dot{M}=4\pi r_{p}^2\rho_p\mathcal{M}(r_p)c_s\label{eqn:mdot_model}
\end{equation}
where the sub-script $p$ refers to evaluating at the outer boundary of the {\sc mesa} model (i.e. the $\tau=2/3$ surface).}

 At early times when the planet is highly inflated ($R_p \sim R_B$), { we find} this rate may formally exceed the so-called ``energy-limited'' rate, the rate at which all stellar flux is converted into kinetic energy in the wind.}{ For these short periods, we cap the mass-loss rate at $10\%$ of the energy-limited rate.} This choice makes little difference to the evolutionary tracks.

{ One also has the freedom in how to implement the mass-loss term in detail, either over a few grids near the photosphere, or spread it across the surface scale heights.}  We have experimented with this and found that provided   the loss term is smooth, small and does not remove significant mass from where the contraction of the atmosphere is taking place, it { does} not matter.

\subsubsection{Implementation of flux boundary condition}

The outer boundary is set   at the $\tau=2/3$ surface to the outgoing radiation. To include stellar irradiation, we adopt the {\sc mesa} $F_*-\Sigma$ implementation \citep{Paxton2013a}.  This consists of depositing a total  irradiation flux of $F_*/4$  uniformly down to a column  density $\Sigma$, where we choose $\Sigma=250$ g cm$^{-2}$  as is appropriate for an opacity of $\kappa_v=4\times10^{-3}$ cm$^2$ g$^{-1}$ to the incoming stellar flux, {\bd the value suggested by \citet{Guillot2010} for sun-like stars}. 

The $F_*-\Sigma$ method of including irradiation in planetary evolution has been shown to provide good agreement to other boundary conditions in \citet{Paxton2013a}, when mass-loss is not dynamically important.  It differs from the usual approach of
 applying a grey or semi-gray approximate solution to the atmosphere \citep[e.g.][]{Guillot2010}, as used by \citet{Owen2013b}.  It is necessary in this problem because energy balance in the upper layers of the atmosphere must be taken into account explicitly, including $P{\rm d}V$ cooling and advection.  However,  for the sake of simplicity, we identify numerical models by their respective $T_{\rm eq}$, not their $F_*$ values.

\subsubsection{Initialization}

We initialize our models as hydrostatic planets, with a radius of $R_B$ ($\alpha=1$), with a chosen envelope mass and core mass. {\bd The planet models were created by building a slightly larger than required planet { (at a higher entropy)} that was then allowed to cool to the required value}.  The core radius is determined assuming a pure rock composition that follows the mass-radius relation from \citet{Fortney2007}. {\bd We include radioactive heating from the core as described in \citet{Owen2013b} but it has negligible impact; moreover, we neglect heating from the core's heat capacity as over the short time-scales of interest the core is unable to transfer heat to the envelope \citep[see][for a detailed discussion]{Lee2014}.}  Specifying these parameters uniquely determines the properties of the planet's atmosphere. We then switch on the hydro-dynamic terms and mass-loss and let the models adjust to a new equilibrium (typically this results in a small reduction in radius). We then evolve the model forward using a time-step that satisfies the convergence criterion described in \citet{Paxton2011} as well as limiting the relative change in envelope mass to $<10^{-3}$ each time-step. We stop the evolution after 3~Myr, at which time mass-loss has practically ceased in all our calculations.  {\bd Given the number of simplifications made in order to perform the calculations, we perform a series of tests in the Appendix to ensure that our results are robust.}

 We consider models with two different initial envelope mass-fractions, 10\% and 30\%, {compatible with those} found by core-accretion models at small separations \citep[e.g.][]{Bodenheimer2014c, Lee2014} and
{have long enough cooling times to remain bloated even at disk dispersal}. We note the true mass fraction, at the time of disk dispersal, is likely determined by the fluid mechanics and the thermodynamics around an embedded planet and is beyond the scope of this work. We study three core masses of 3, 5 \& 10 M$_\oplus$ and two equilibrium temperatures of 500 \& 900~K. 

\subsection{The Results}
\label{subsec:results}

\begin{figure}
\centering
\includegraphics[width=1.05\columnwidth]{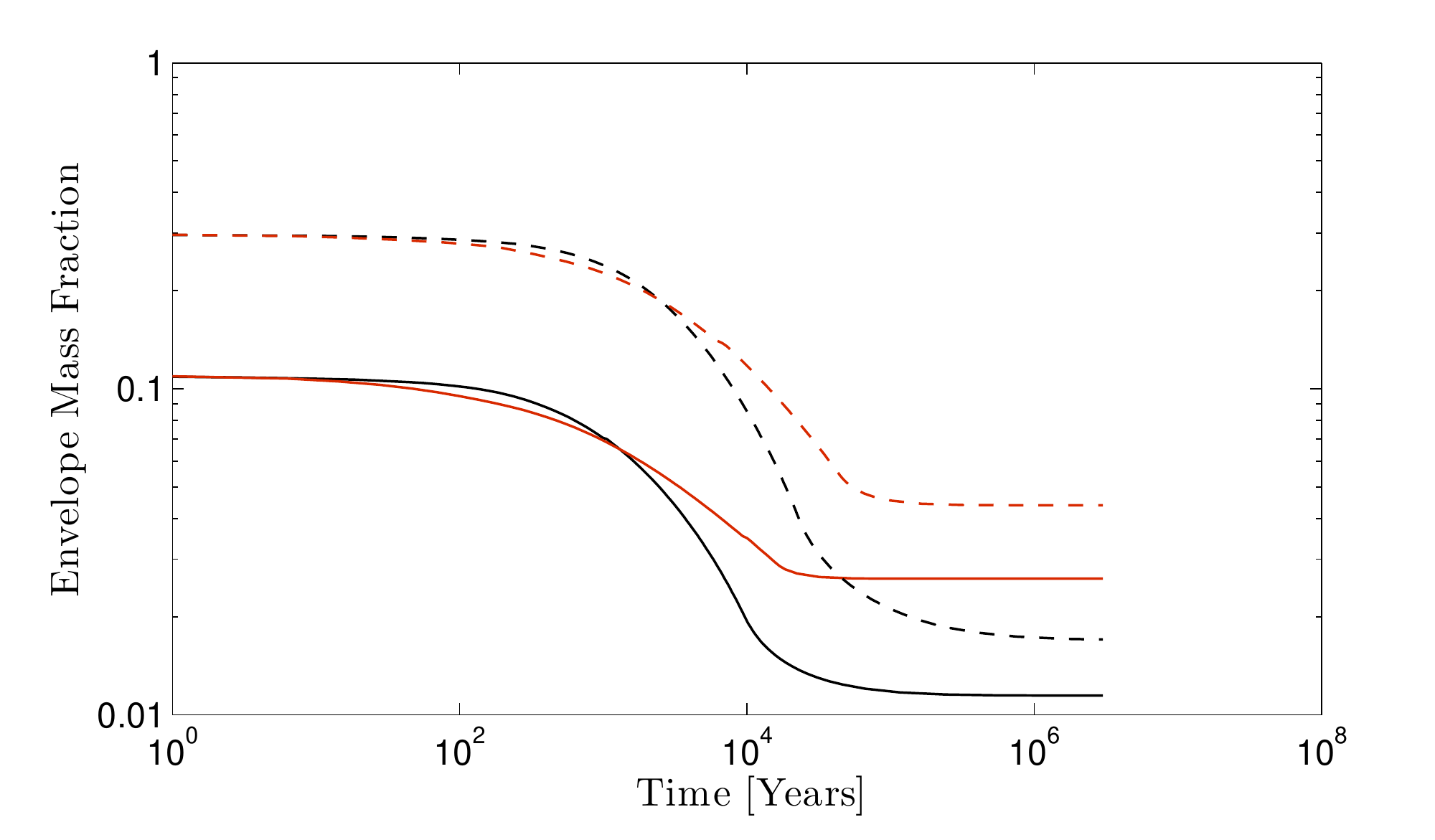}
\caption{Mass evolution  for models with a 5~M$_\oplus$ core. The red lines are for models with an equilibrium temperature of 500~K while the black lines  are for 900~K.  The solid and dashed lines stand for models with initial envelope fractions of 10 and 30\% respectively. All models start with a radius $R_p = R_B$. }\label{fig:5Me_evolve}
\end{figure}

\begin{figure}
\centering
\includegraphics[width=1.05\columnwidth]{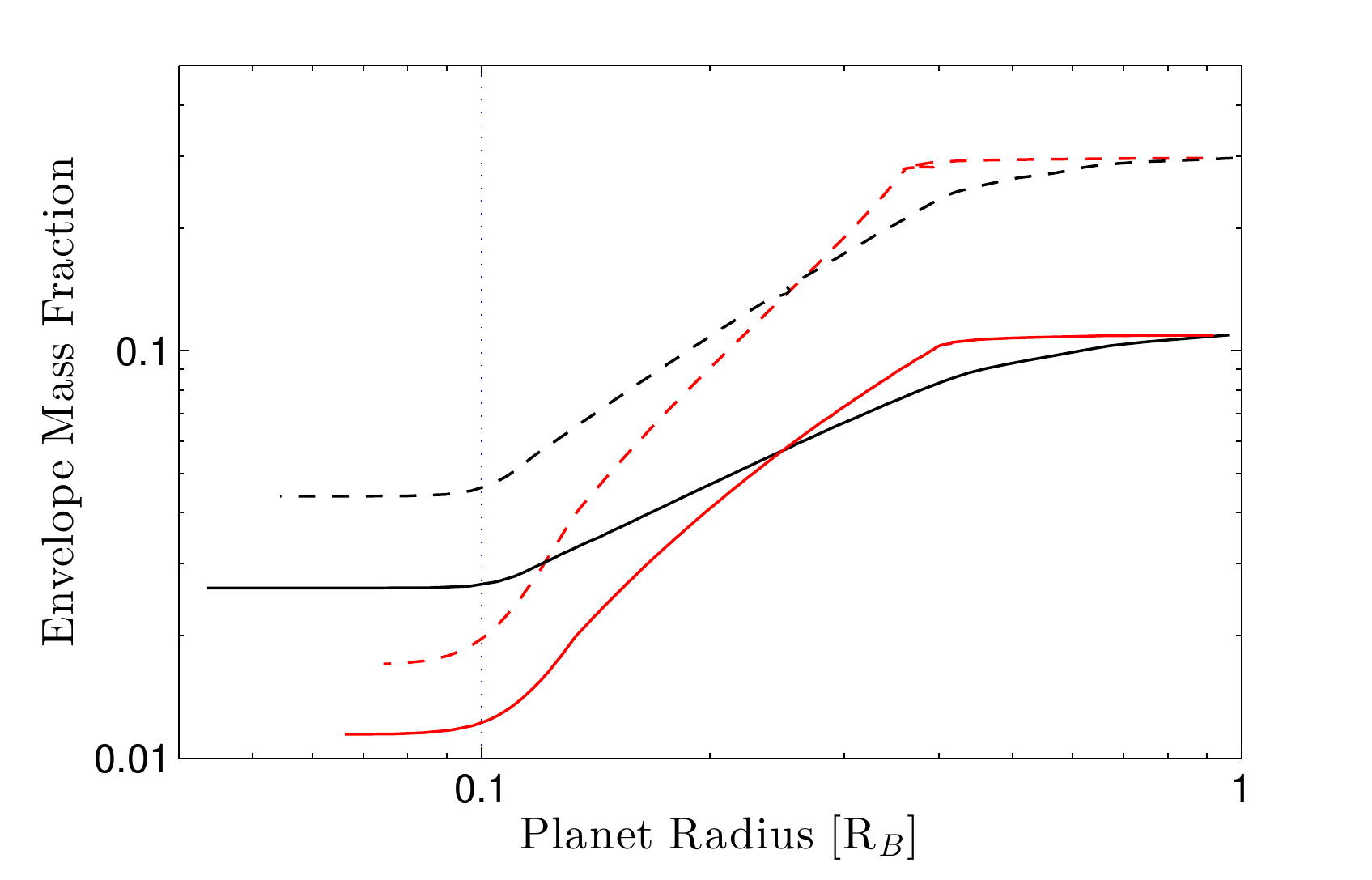}
\caption{Evolution of the same models in Fig.~\ref{fig:5Me_evolve}, shown in the radius-mass plane, where the planet's radius is scaled to the Bondi radius. A radius of 0.1~$R_B$ is shown as the thin dotted line. }\label{fig:5Me_rad}
\end{figure} 

As is shown in Fig. \ref{fig:5Me_evolve}, the evolution of all models is similar: the planet's atmospheres contract, lose mass and cool on a time-scale of order $10^5$~years.   The internal contraction provides energy to lift up the outer envelopes.  The mass-loss drops off exponentially with planet size as discussed in Section~3.1, and after a few $10^5$ yrs, when the planet's radius reaches $\sim 0.1R_B$ (Fig. \ref{fig:5Me_rad}), mass-loss ceases entirely  and the planet cools down normally. At this point, the final envelope mass is $\sim 10\%$ of the initial one, consistent with the estimate in Eq. \refnew{eq:mfinal}.

\begin{figure} \centering \includegraphics[width=\columnwidth]{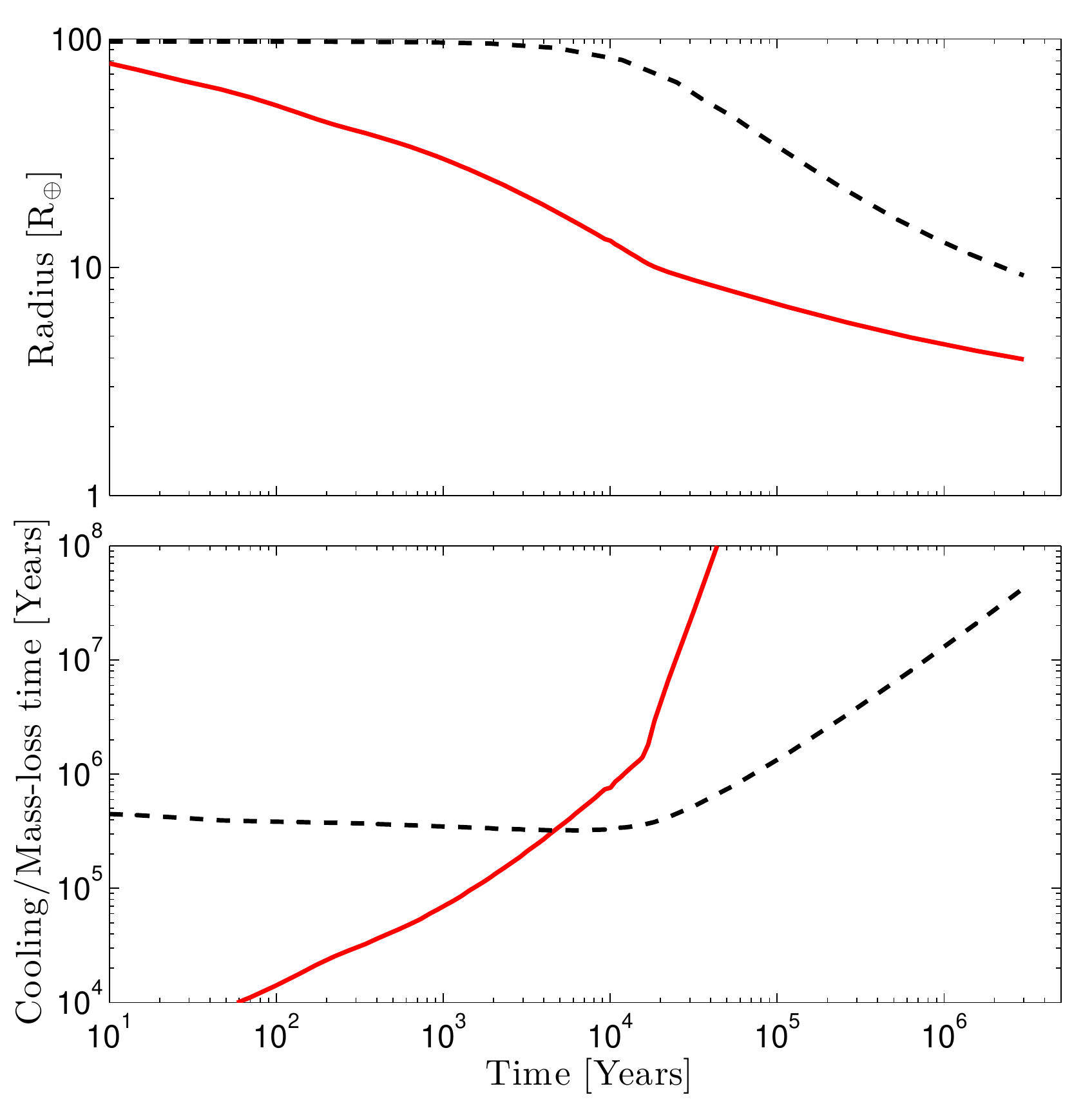} \caption{ Radius and time-scale evolution for a 5~M$_\oplus$ core  and an initial envelope mass fraction  of 10\% , at an equilibrium temperature of 500~K (the solid red line in Fig. \ref{fig:5Me_evolve}).   The top panel shows its radius evolution with the dashed curve representing a similar model where there is no mass-loss. The bottom two panels shows how the Kelvin-Helmholtz time-scale (dot-dashed) and  the mass-loss time-scale (solid) evolve as a function of time and envelope mass fraction.  By the time the mass-loss ceases ($\sim 1$ Myrs), the planet looks like one that has cooled for $\sim 50$ Myrs. }\label{fig:time-scales}
\end{figure}

 We focus on one model to gain more insight. Fig. \ref{fig:time-scales} presents the evolution of a planet with a $5 M_\oplus$ core, an initial mass fraction of $10\%$ and at an equilibrium temperature of $500\K$ (the red solid line in Fig. \ref{fig:5Me_evolve}).
 Compared to a similar model without an imposed mass-loss, the contraction of our model is much more rapid, indicating that the energy for the mass-loss indeed obtains from internal gravitational contraction. 
The mass-loss time-scale (inverse of the mass-loss rate) starts low but sky-rockets to astronomically long time-scale as the planet's radius shrinks. The Kelvin-Helmholtz time, defined as the time-scale to radiate the envelope's binding energy at the current luminosity,\footnote{We pick this to be the maximum luminosity in the envelope, rather than the surface luminosity as the latter can sometimes be negative (Fig. \ref{fig:snapshot}).}
lies always above the system age, again indicating that the binding energy is not primarily lost by radiative/convective transport, but by advection. By the time mass-loss has stopped ($\sim 1$ Myrs), the planet has a Kelvin-Helmholtz time of $\sim 50$ Myrs. It has ``aged'' prematurely.

\begin{figure*} \includegraphics[width=0.9\textwidth]{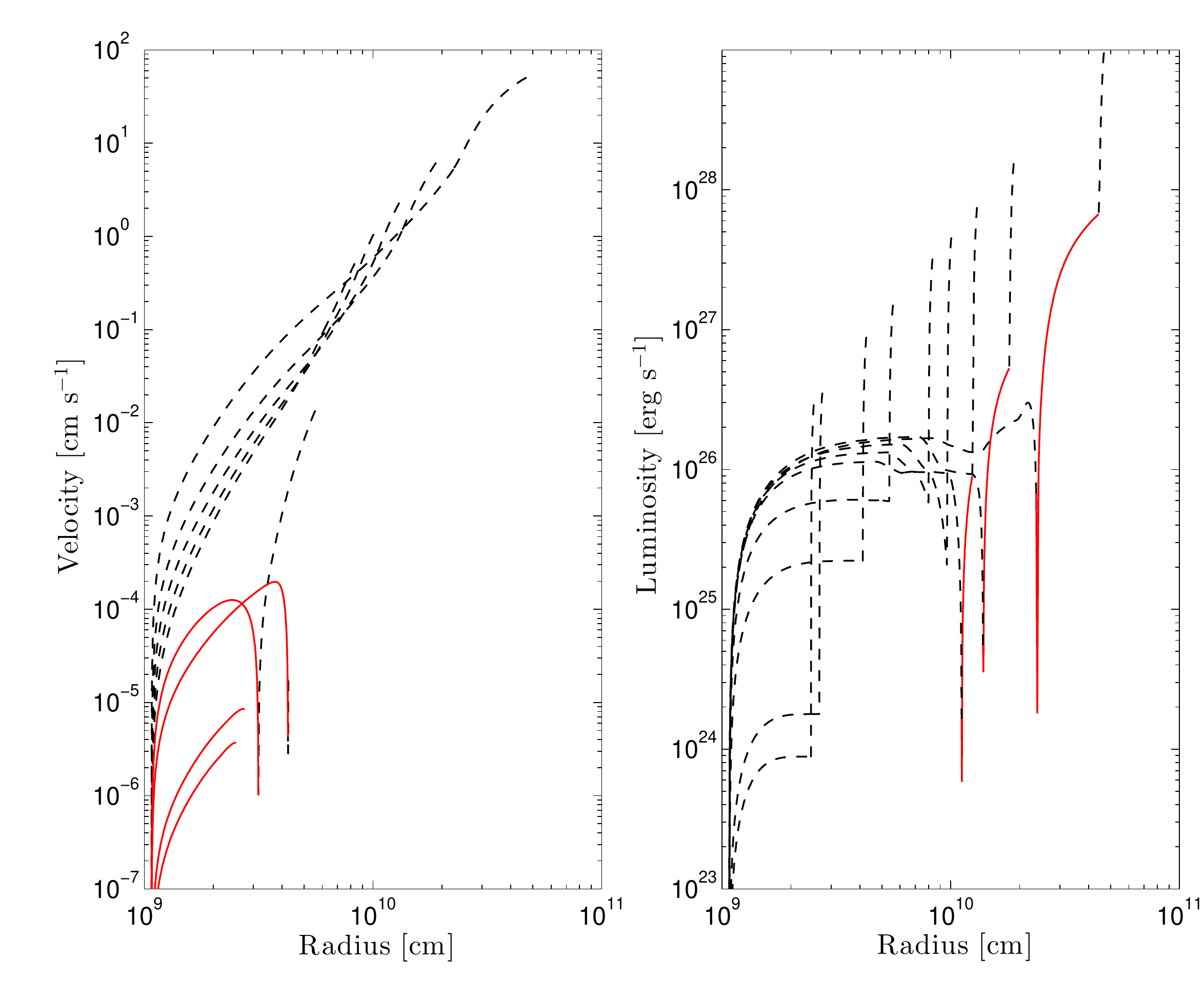}
 \caption{Snapshots of the velocity and luminosity profiles as a function of radius for a model with a $5 M_\oplus$ core, an initial mass fraction of $10\%$ and at an equilibrium temperature of $500\K$ (the model shown in Fig. \ref{fig:time-scales}).  From top to bottom, the following times are shown: 10$^2$, 10$^3$, 10$^4$, $3\times10^4$, $6\times10^4$, 10$^5$, $3\times10^5$, 10$^6$ \& $3\times10^6$ years.  Solid lines represent negative values, while dashed lines represent postive values.  A negative luminosity represents an inwardly directed radiative flux. The large luminosity spike at the surface  arises as we are explicitly adding the heating from stellar irradiation and is representative of  the re-radiated stellar flux. {\bd 
     The change in sign close to the planet's surface is abrubt and unresolved, as a result of our simplified $F_*-\Sigma$ method for including stellar irradiation. However, as discussed in the appendix this simplified treatment does not impact our calculations.}}\label{fig:snapshot} \end{figure*}

 We present snapshots of the  fluid velocity  and luminosity profiles for the same model in Fig. \ref{fig:snapshot}.  Initially the planet's envelope is expanding so fast, it completely absorbs the planet's internal luminosity  and converts it into $P$d$V$ work, resulting in  negative luminosity in the surface layers. This layer  reaches much deeper than the penetration depth of the stellar irradiation, again proving that  the main source of energy for the mass-loss is the planet atmosphere's internal luminosity.  By losing mass, the planet  is then able to cool and contract significantly such that it shuts off the mass loss.  Within 1~Myr, the planet's luminosity has dropped by a factor of 100, and its envelope now radially contracts (as opposed to outflow).

\begin{figure}
\centering
\includegraphics[width=\columnwidth]{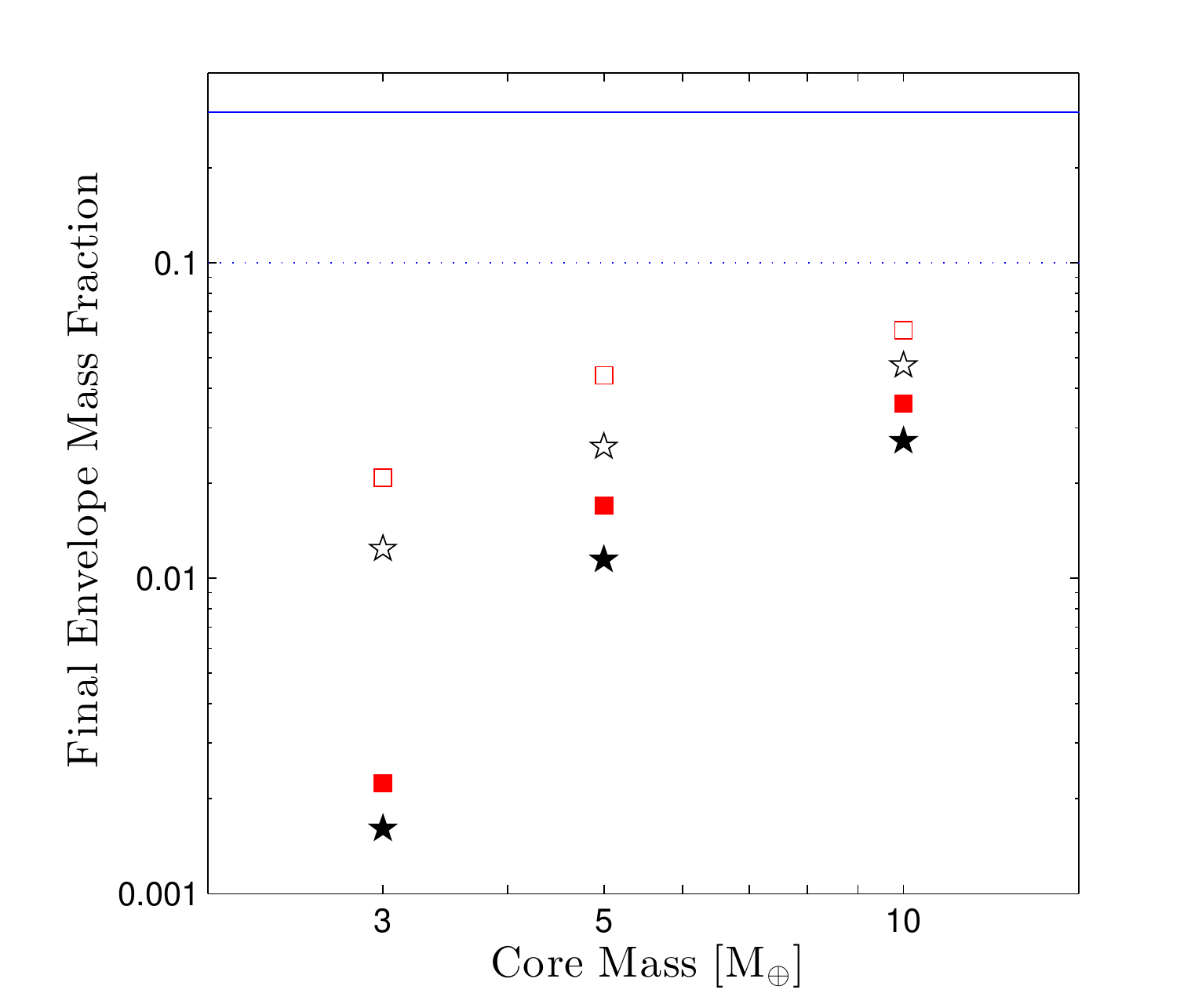}
\caption{Final envelope mass as a function of core mass. Red symbols have an equilibrium temperature of 500~K, whereas black are 900~K. Open symbols initially had 30\% envelope fraction (shown as the thin solid blue line) and the filled symbols initially had 10\% envelope fraction (shown as the thin dotted blue line). { Assumed to start at a radius of $R_B$, all models lose substantial amounts of envelopes, with the effect being more drastic for planets with a lower core mass, a stronger irradiation and a less massive envelope. } }\label{fig:final_mass}
\end{figure}

\begin{figure}
\centering
\includegraphics[width=\columnwidth]{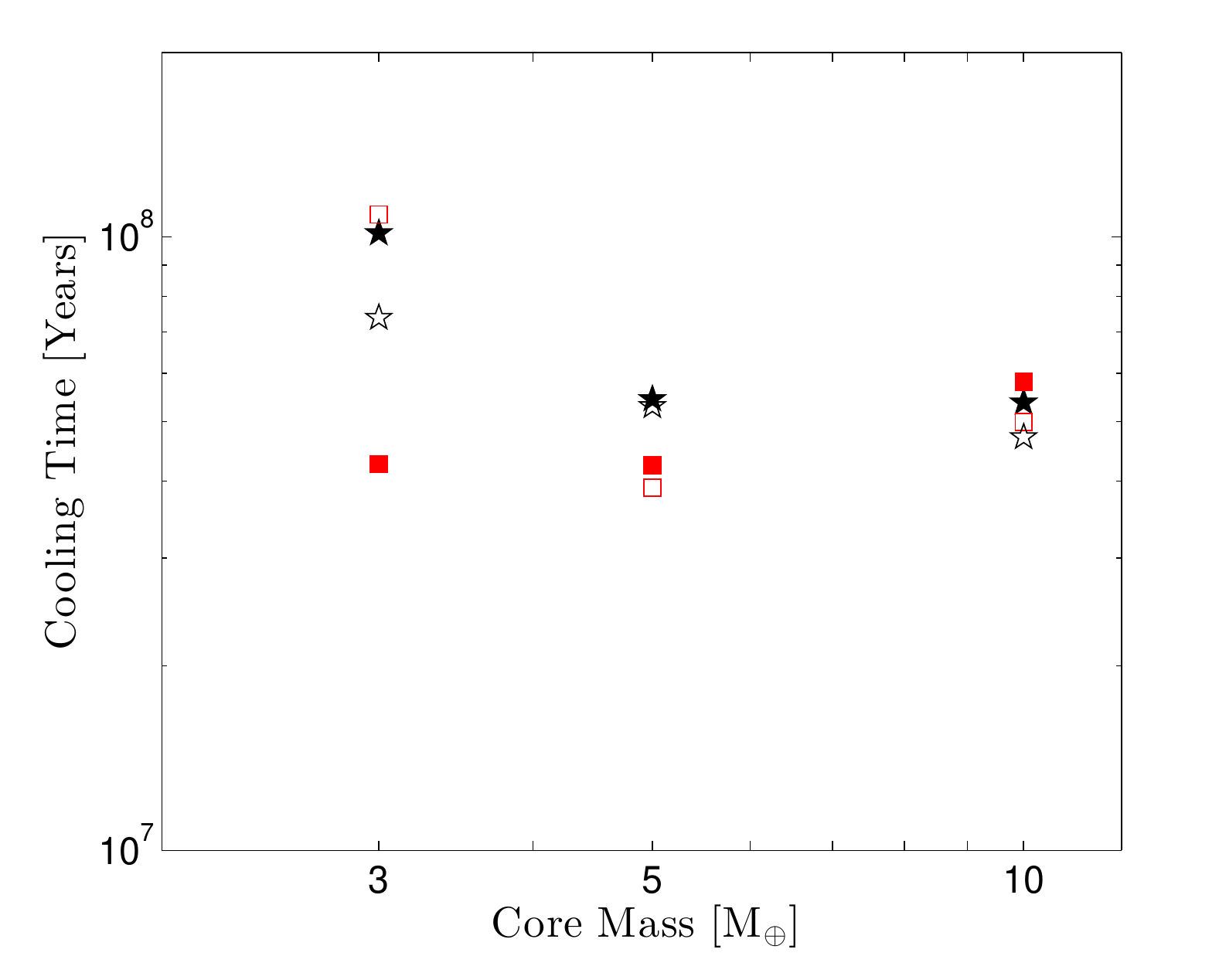}
\caption{Same as Fig.~\ref{fig:final_mass}, but for  the cooling time 
 at the end of $3$ Myrs. { The ``boil-off'' removes internal energy from these planets, allowing them to cool off much faster than their ages indicate.}}\label{fig:final_tcool}
\end{figure}

\begin{figure} \centering \includegraphics[width=\columnwidth]{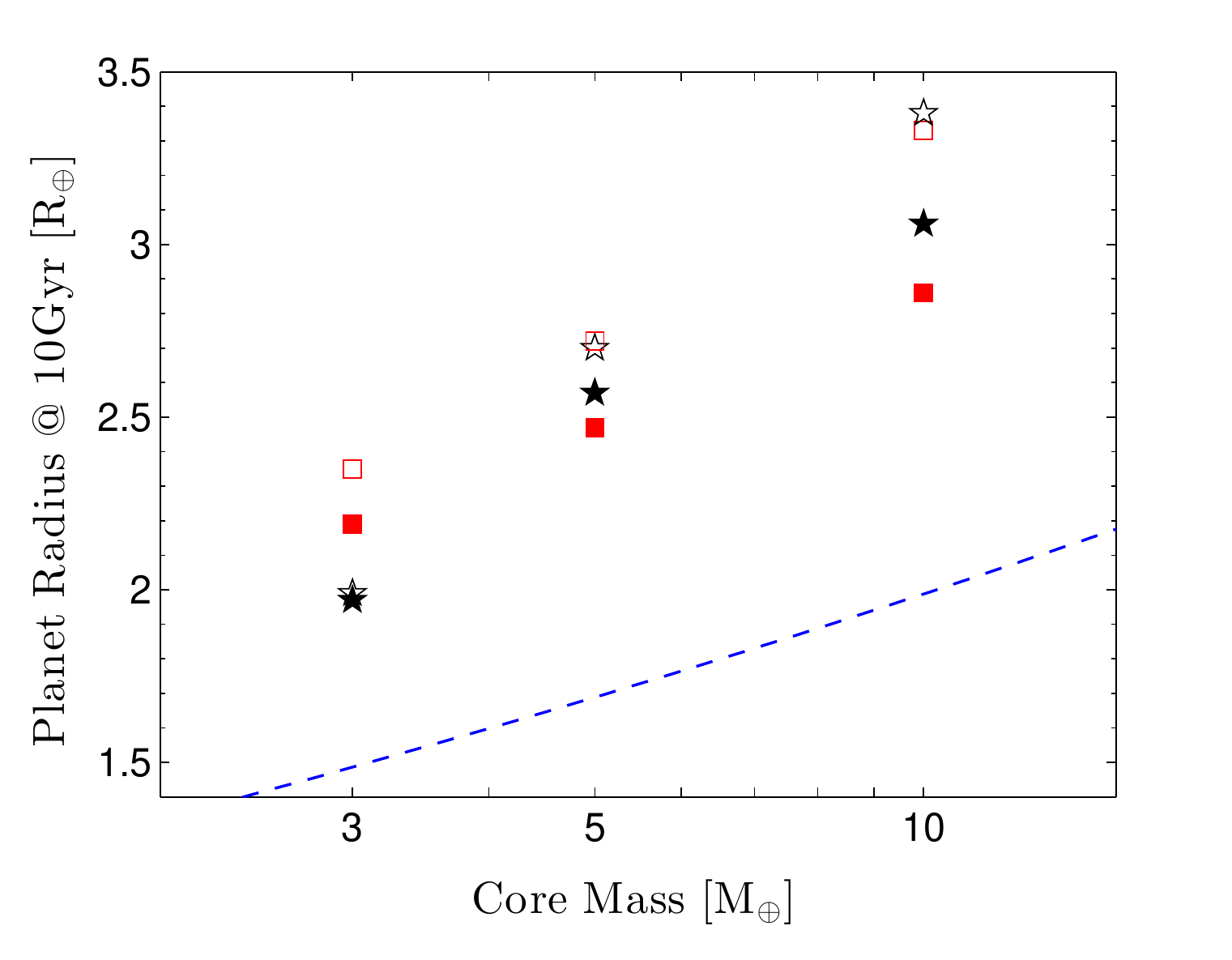} \caption{Same as Fig.~\ref{fig:final_mass}, but for the final planet radius { after} $10$ Gyrs { of cooling contraction}.  For planets less massive than $10 M_\odot$, the final radii tend to cluster around $2.5 R_\oplus$, the radius observed for a large number of {\it Kepler} planets.    Here, the dashed line indicates the radius of the planetary core. }\label{fig:final_radius}
\end{figure}

Finally, in Figs.~\ref{fig:final_mass} \& \ref{fig:final_tcool} we present composite results  for the envelope mass and the cooling time  after $3\times10^6$ years  of evolution.  We see that models receiving a stronger stellar insolation lose more mass -- roughly by a factor of 2 between the 900~K and the 500~K models. {\bd  We also see that less massive planets have a harder time retaining their envelopes. {\bdd This difference is expected from the expression given in Eqn. (17), where stronger  irradiation and lower core masses lead to slightly more mass-loss, although we note that since the this effect is small and the planets cool significantly, energy input from stellar irradiation is a sub-dominant effect.} In some cases, envelopes around the lowest mass planets (3 M$_\oplus$) can be blasted away so completely, they appear as ``naked'' cores.  Furthermore,  by the end of $3$ Myrs, all planets have a cooling time in the range  of $4\times10^7-10^8$~years, approximately $\sim20$ times longer than one would have predicted if the planets cooled purely by gravitational contraction.

\section{Discussion}\label{sec:discuss}

 We discuss how our {new process}  relates to previous works on mass-loss of planetary envelopes, and what our work implies for the observed planetary properties. {\bd We connect this process with the first stage of mass-loss, that occurs while the disk is dispersing \citep{Ikoma2012} and with the later stage driven by EUV/X-ray photoevaporation.}

\subsection{Relation to \citet{Ikoma2012}}
 \citet{Ikoma2012} considered the evolution of an embedded low-mass planet as the  background disk density and pressure slowly declined.  They found that the planetary atmosphere was eroded, more extremely so if an additional heat source keeps the planet puffy.  {For the latter, t}hey invoked the cooling luminosity from the solid core. However, \citet{Lee2014} argued they may have adopted a cooling luminosity that is too large -- with an assumed cooling time of $\sim 10^5$ yrs for a core the size of the Earth -- \x{\citet{Lee2014} demonstrated they} {this may have} grossly overestimated the transport ability of either heat conduction or mantle convection.

\citet{Ikoma2012} is concerned with the mass-loss during the disk dispersal (with a time-scale $\sim 10^5$ yrs), while we are interested in what happens afterwards.  So the process they investigated is a natural predecessor to the one studied here, and it provides the initial input for our model.

 Physically, their mechanism is similar to ours, except in details, at least in the case of no additional heat source.  By reducing external pressure confinement gradually, \citet{Ikoma2012} observed a mass-loss that likely occurs through a sub-sonic breeze.\footnote{However, this process is not explicitly modelled in their work and as a result, their mass-loss rate may not be physically self-consistent.}
{ In this work, we are concerned with a transonic wind.} Both flows { are similar in that they both} draw energy from the gravitational contraction (when core cooling is not important). { Above} the photosphere, our wind is propelled by stellar heating, while their mass-loss is powered by the pressure differential between the atmosphere and the background.

How important is this early episode of mass-loss? or, in other words, when does the breeze stop and the wind start? We speculate that the transition may occur well before the disk becomes so optically thin  that the outflow can be heated by the star. The infra-red glow from the background disk may be already sufficient to keep the outflow roughly isothermal.
{\bd Moreover, the decay of the disk's confining pressure becomes irrelevant once the ram pressure in the outflow is strong enough. This occurs before} the gas mean free-path becomes as large as the planetary radius, a threshold set by \citet{Ikoma2012} for the termination of their simulations. However, in order to model this properly, one need to improve upon the modelling in \citet{Ikoma2012}, by including the hydrodynamics terms explicitly in the evolution equations, { and to include the breeze solution explicitly. {\bd Ultimately a single simulation that follows the protoplanet all the way through disk dispersal and boil-off is required to assess the relative importance of the processes in shaping the planet population.}

 {\bd We note that in some of our numerical calculations the mass-loss time-scale is similar to the disk clearing time scales and as such our results will be sensitive to when this cross over takes place. Specifically, if disc pressure drops quicker than the ``breeze'' solutions can transfer mass sufficiently to maintain equilibrium, then the planet will launch a Parker wind early; however, if the disk pressure drops slowly enough,  the ``breeze'' can remove sufficient material \x{allowing this first phase of mass-loss to proceed until longer, perhaps meaning the boil-off phase is not as important as described here.} {\bd to make the wind phase (boil-off) less important. We note that the breeze mass-loss rates are always lower than those driven by the Parker wind, thus we suspect the ``boil-off'' phase will dominate the total mass-loss.}
}


\x{During this disk removal stage the proto-planetary atmosphere is likely to interact with the dispersing disk transferring mass, probably through a sub-sonic breeze.  The envelope mass at the end of this process (start of the models presented here) is likely to depend on the details of the disk removal process, it is still fast, such that once the proto-planet is exposed to vacuum boundary conditions it will be pre-deposed to mass-loss as discussed in Section~\ref{sec:setup}. However, should disk dispersal be much slower than currently thought, or the opacity in the proto-planetary envelope be sufficiently different from what is assumed then the envelope might be able to cool and contract during disk dispersal and not be predisposed to mass-loss once it is exposed to vacuum boundary conditions. }

\subsection{Relation to EUV/X-ray driven evaporation}

More mass-loss can occur at the end of our simulations, after the planet has contracted to a radius of $R_p \leq 0.1 R_B$. The ionizing radiation (EUV/X-ray, as opposed to continuum radiation considered here) from the star can elevate the temperature in the upper layers to $\sim 5000-10^4\K$, as opposed to the blackbody temperature considered here \citep{Owen2012c}. {The corresponding sound speed can approach the escape velocity at the photosphere, even for a more compact planet. This then drives a photoevaporative outflow.} { This has been shown to be particularly significant for {\it Kepler} planets inward of $\sim 0.2$ AU from the host stars and can explain the smaller sizes of these close-in planets} \citep{Lopez2013,Owen2013b}. The atmosphere of these planets can be completely removed, consistent with the high bulk densities observed in objects like CoRoT-7b \citep{Hatzes2011}, Kepler-10b \citep{Batalha2011}, Kepler-36b \citep{Carter2012} and others \citep{Weiss2014a}.

\citet{Owen2013b} demonstrated that this mode of mass-loss is most significant in the first $\sim 100$ Myrs, mostly because this is when the host star is more chromospherically active (and hence stronger ionizing radiation), but also because this is when the planet is more extended (with a smaller surface escape velocity). Short of knowing the initial radii for the planets, they explored models with a range of initial cooling times (from $3$ Mys to $100$ Myrs), where a shorter cooling time corresponds to models with a more extended photosphere. They concluded that models with the shortest cooling time lose the most mass, with the difference being more significant for lower mass envelopes (Fig. 4 of \citealt{Owen2013b}).

This provides the context for our results in Figs. \ref{fig:final_mass}-\ref{fig:final_tcool}. They quantify the initial conditions for the later photoevaporative flow, just like results from a study like \citet{Ikoma2012} can do for our work here. The early episode of { mass-loss}, propelled by continuum stellar radiation and powered by gravitational contraction of the planet, cools the planet to a smaller radii, affecting the outcome of later photoevaporation. As such, we argue that a coherent study where one follows the evolution of the planet envelope through these different stages, instead of studying them independently, would be needed to predict the final envelope mass one observes.

\subsection{Applications to {\it Kepler} Planets}
Despite being only one of the three possible steps in mass removal, the ``boil-off'' process discussed here may explain some of the observed features in {\it Kepler} planets.

It may {\bd help} explain the dramatic deficit of planets with sizes above $\sim 2.5 R_\oplus$ in the {\it Kepler} catalogue. This radius corresponds to an envelope mass of $\sim 1\%$ for planets that have cooled for a few Gyrs. Limited by the transit technique and the mission duration, this catalogue mostly contain planets inward of $\sim 0.5$ AU ($T_{\rm eq} \approx 400\K$).  For these planets, Fig. \ref{fig:final_mass} shows that, even if they were started with much more massive envelopes ($10-30\%$), the ``boil-off'' would have left them with much punier envelopes (of order $1\%$) after a few Myrs. Moreover, the final envelope mass, at the same location, is found to scale roughly with the core mass. As a more massive core exerts a stronger gravity on the envelope and compresses it more, this correlation leads to a further convergence in final planet radii, as is shown in Fig. \ref{fig:final_radius}. 


In contrast, planets outside this distance range will experience less mass-loss (see Fig. \ref{fig:5Me_evolve}) and they may retain most of their primordial envelopes.  In other words, we predict that there are relatively more neptunes, with envelope mass fractions of tens of percent, outside the $\gtrsim 1$~AU range. This could be tested by future transit missions. 

{\bd We can contrast these predictions of the boil-off against those from EUV/X-ray photoevaporation \citep{Owen2013b}. The latter effect is mostly limited to separation $\lesssim 0.2$ AU. If the latter acts alone, one would expect an abundance of Neptunes outside $0.2$ AU, closer than our above prediction for the ``boil-off''.  This difference may already be discernible in the {\it Kepler} data. However, detailed comparison between theory and observation needs to take into account of the planet mass, which is largely unknown at the moment, especially for planets outward of $0.2$ AU.}


{\bd Are there Neptunes within 1AU? We find that {\it Kepler} data show} a smattering of neptune-like planets {\bd within this distance}. How could these planets have escaped the ``boil-off''? We speculate that their cores are more massive than $\gtrsim 10$ M$_\oplus$ (see Fig. \ref{fig:final_mass}) and therefore can hold on to their envelopes better. This conjecture is supported by { the general mass-radius relationship of $M\propto R$, as is obtained from} TTV and RV mass measurements \citep{Wu2013,Hadden2014,Weiss2014a}.  \x{ where one observes a mass-radius relation of the form $M \propto R$, i.e.,} {The fact that} larger planets do tend to have more massive cores {is consistent with the ``boil-off'' process.}  Conversely speaking, the ``boil-off'' {\it helps} explain the observed mass-radius relation.

{ However, such an explanation is incompatible with the presence of a few Neptune-like planets which have unusually low densities \citep[Kepler-51,][]{Masuda2014} \citep[Kepler-79,][]{Jontof-Hutter2014}; \citep[Kepler-87,][]{Ofir2014}. 
These masses are typically of order $2-5 M_\oplus$, within the range where we predict the ``boil-off'' should have carried off most of their envelopes. They may be counter evidence for our theory, but we hope their mass determinations (all TTV inferences) can be confirmed by radial velocity studies. Furthermore, since they do not appear a dominant population (through the radius distribution) and they may have a different origin \citep{Lee2015b}.} 

Finally, \citet{Rogers2014a} suggest that all planets with radii $\lesssim 1.6$~R$_\oplus$ are predominantly rocky. A $1.6$~R$_\oplus$ rock planet is roughly a 4~M$_\oplus$ \citep{Fortney2007}. We note that this process may play a role in shaping this result. However, we need to be cautious as one needs to perform a more detailed study that factors into the separation distribution, as normal evaporation \citep[e.g.][]{Owen2013b} is able create this as well. Thus, in order to untangle the true origin of the solid to H/He transition -- which may turn out to be a separation dependant statement -- further modelling is required.

\x{\subsection{Issues and future directions}}

\section{Conclusion}

We now know  that there are three stages of envelope mass-loss: while the planet is embedded in the disk; soon after the disk disperses; and while the planet is contracting, with typical durations of $\sim 10^5$ Myrs, $10^5$ Myrs and $10^8$ Myrs, respectively. These processes combine to sculpt the final radius and mass distributions of the close-in low-mass planets.

 We have presented and investigated the second stage, dubbed the ``boil-off'' phase, whereby an irradiated, inflated ($R_p \sim R_B$) low-mass planet loses its envelope and contracts quickly.  Our investigation can be summarized as follows:

\begin{enumerate}

\item   We argue that many low-mass planets should have photospheric radii $R_p \sim R_B$ by the time disk disperses, 
due to their naturally long Kelvin-Helmholtz time-scales, {\bd and perhaps aided by the small amount of heat input by planetesimal accretion.}. As such, they are subject to the ``boil-off''.   The primary energy source for the mass-loss is the binding energy of the planet itself, with  the continuum stellar irradiation acting as a ``catalyst'' for the mass-loss. 

\item  Due to the characteristics of {an isothermal wind}, this ``boil-off'' will cease when the planet has contracted to within $0.1 R_B$.

\item  {\bd Order of magnitude} { energetic considerations} show that by this time, the envelope mass has reduced 
to $\sim 0.2/\alpha$ of its initial mass, where $\alpha$ is the initial radius in unit of $R_B$.

\item 
We find that this process leads to extreme cooling of the planet.   Within a Myrs, the planet have cooled down to a state that is only reached after $\sim 100$ Myrs of evolution without mass-loss.

\item Relatively speaking, mass-loss is more severe  {for} planets with lower core masses,  and {for} planets subject to stronger stellar irradiation. Planets more massive than $10 M_\oplus$ should be {\bd less affected by} this process, so do planets that have accreted so much gas they have undergone run-away gas accretion.

\item  This process could remove the massive envelopes of neptune-like planets that are inward of $\sim 0.5$AU, explaining the observed steep fall-off in planet numbers
{above} a size of $2.5 R_\oplus$, {\bd provided planet formation results in initial envelope mass fractions of order tens of percent, as is suggested by recent simulations and basic time-scale arguments}, {\bdd \citep[e.g.][]{Lee2014,Lee2015b}}. \x{Accordingly,} A corollary  {\bd is that} we expect neptunes to be more common at larger separations.

\item  This process may also be responsible for the observed mass-radius relation ($M \propto R$). More massive planets can retain more of their envelopes, leading to larger sizes. {\bd However, EUV/X-ray photoevaporation can also lead to the same trend. Work is needed to identify the relative importances of these processes.}

\end{enumerate}

 To predict the final envelope mass for planets, one need to model the formation, the disk dispersal, the ``boil-off'' and the subsequent photoevaporation process, a task that may now be within reach. 


\acknowledgements We thank the anonymous referee for a thorough and insightful report. JEO acknowledges support by NASA through Hubble Fellowship grant HST-HF2-51346.001-A awarded by the Space Telescope Science Institute, which is operated by the Association of Universities for Research in Astronomy, Inc., for NASA, under contract NAS 5-26555. YW acknowledges financial support from NSERC. We are grateful to Eugene Chiang for fruitful discussions.  

\bibliographystyle{apj}
\bibliography{library}

\appendix

\section{Tests of Numerical Approach}
In order to model the ``boil-off'' phase with the {\sc mesa} stellar/planetary evolution code we have made a number of simplifying assumptions in order to model the radiative transfer and mass-loss. Therefore, it is important to check that these approximations are not driving our results and our results are robust to the exact tuning parameters chosen in {\sc mesa}. When including mass-loss from the boil-off process in the calculations we assume that the flow is isothermal and can be approximated by a Parker wind which we do not attempt to explicitly model. Instead the outer boundary for our models is the photosphere to the outgoing IR radiation and we apply the mass-loss as a sink term at the top of the envelope. We can check the outer layers of our model (below the photosphere) are quasi-isothermal and close to the equilibrium temperature indicating that our choice of a Parker-wind is an appropriate first step. If, for example $P{\rm d}V$ work of the expanding upper atmosphere (below the photosphere) could not be compensated for by the radiative heating then the atmosphere would significantly cool below the equilibrium temperature, being closer to adiabatic, then a Parker wind would not be an appropriate model.  

\begin{figure}
\centering
\includegraphics[width=0.5\textwidth]{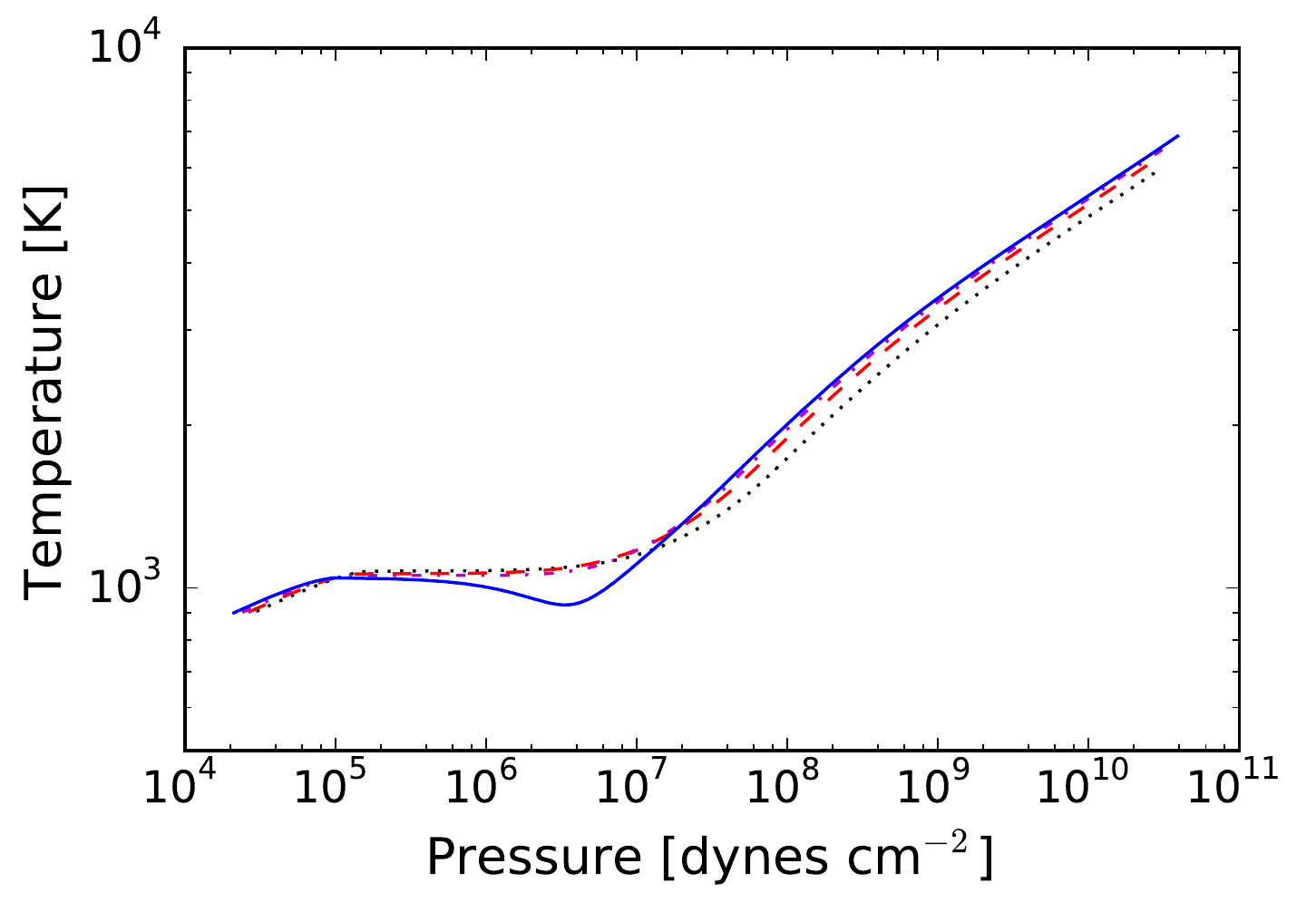}
\caption{Pressure temperature profiles for the calculation of a planet with an initial envelope mass fraction of 30\%, core mass of 5~M$_\oplus$ at an equilibrium temperature of 900~K. They are shown at times of $3.7\times10^{4}$ (solid), $8.3\times10^{4}$ (dot-dashed), $1.8\times10^5$ (dashed) and $4.4\times10^{5}$ (solid) years.}\label{fig:P_T}
\end{figure}

Snapshots of the pressure-temperature profile are shown for an evolving planet in Figure~\ref{fig:P_T}, which show the outer layers of the envelope are quasi-isothermal during the evolution close to the equilibrium temperature of 900~K.

As discussed in Section~4, the mass-loss is included as a sub-step in the evolution of the planet's envelope where mass is removed from the upper layers. As noted above, occasionally at early times the mass-loss rate given by the Parker wind occasionally exceeded the ``energy-limited'' rate and we included an efficiency factor of $10\%$ to correct this. In Figure~\ref{fig:eff_compare} we show the evolution of the envelope mass fraction for the model with a 5~M$_\oplus$ core and initial envelope mass fraction of $30\%$ where this efficiency is varied with values of 10\% (solid) and 100\% (dot-dashed), while this obviously effects the evolution at early times, the evolution is identical at late times with the final envelope mass fraction are within $<0.3$\% across all the simulations.

\begin{figure}
\centering
\includegraphics[width=0.5\textwidth]{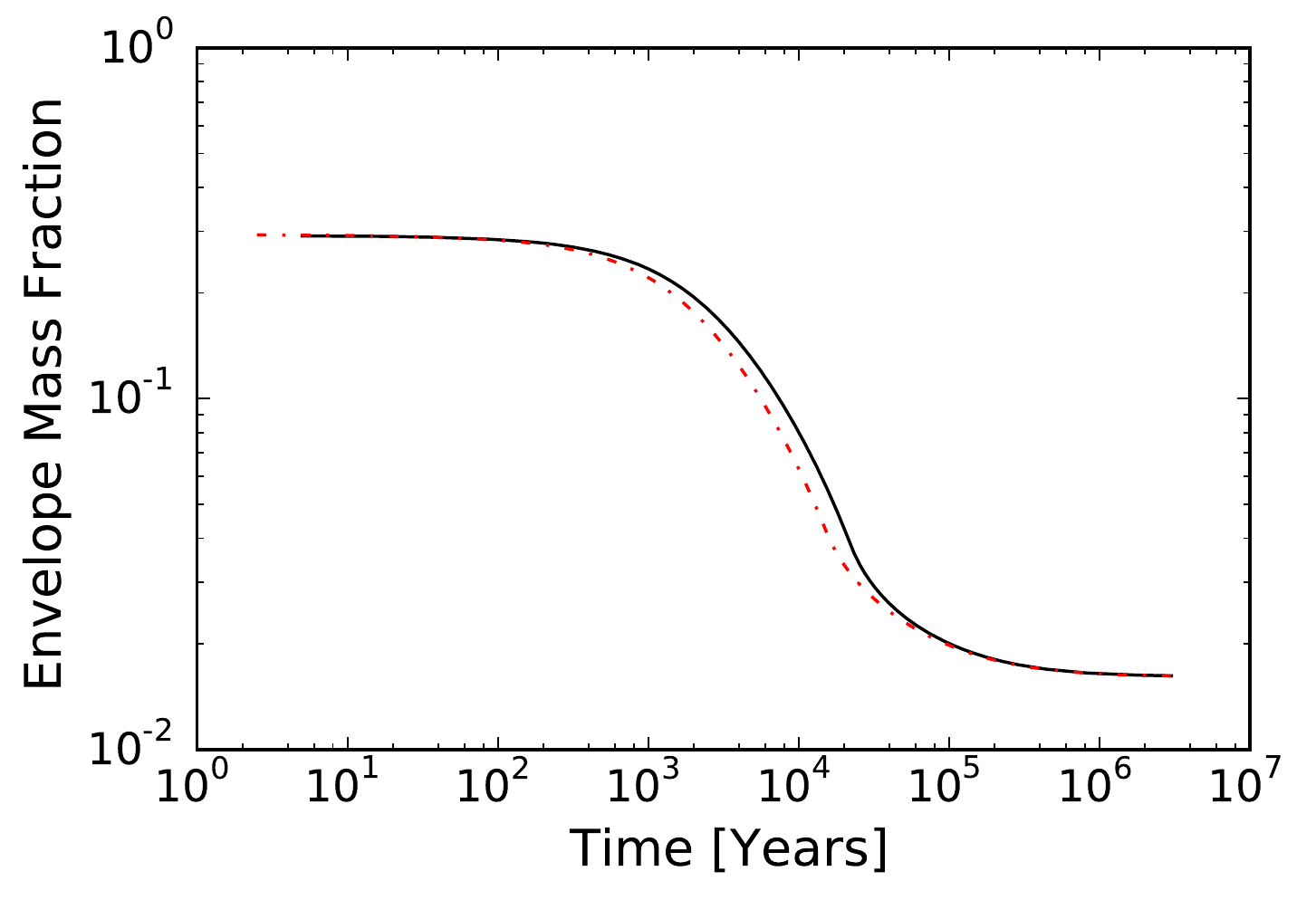}
\caption{Evolution of the envelope mass fraction for the model with a 5~M$_\oplus$ core and initial envelope mass fraction of $30\%$ at an equilibrium temperature of 900~K. The different lines show cases where the mass-loss rates is capped to different values of the ``energy-limited'' rate at early times: 10\% (solid - the standard value used in our work) and 100\% (dot-dashed). The evolution at early times is different, but at late times the evolution is identical and the final envelope mass fractions found are consistent to within $<0.3\%$. Therefore, we are confident the efficiency choice is not driving our results.}\label{fig:eff_compare}
\end{figure}

We can also check that the amount of the atmosphere we are smoothing the mass-loss over does not effect our results. In {\sc mesa} the range of the envelope that material is removed every sub-step is controlled by the 
\verb|min_q_for_k| parameters that set the mass fraction of the upper envelope over which material is removed. The standard value is that any mass is removed from the top (by mass fraction) $<0.5\%$ of the atmosphere, which is used in all our previous calculations. In Figure~\ref{fig:env_remove} we show the evolution of the envelope mass fraction for  the model with a 5~M$_\oplus$ core and initial envelope mass fraction of $30\%$ where the envelope mass fraction material is removed from is $<0.5\%$ (solid), $< 1\%$ (points), $<2\%$ (dashed) is shown. 
\begin{figure}
\centering
\includegraphics[width=0.5\textwidth]{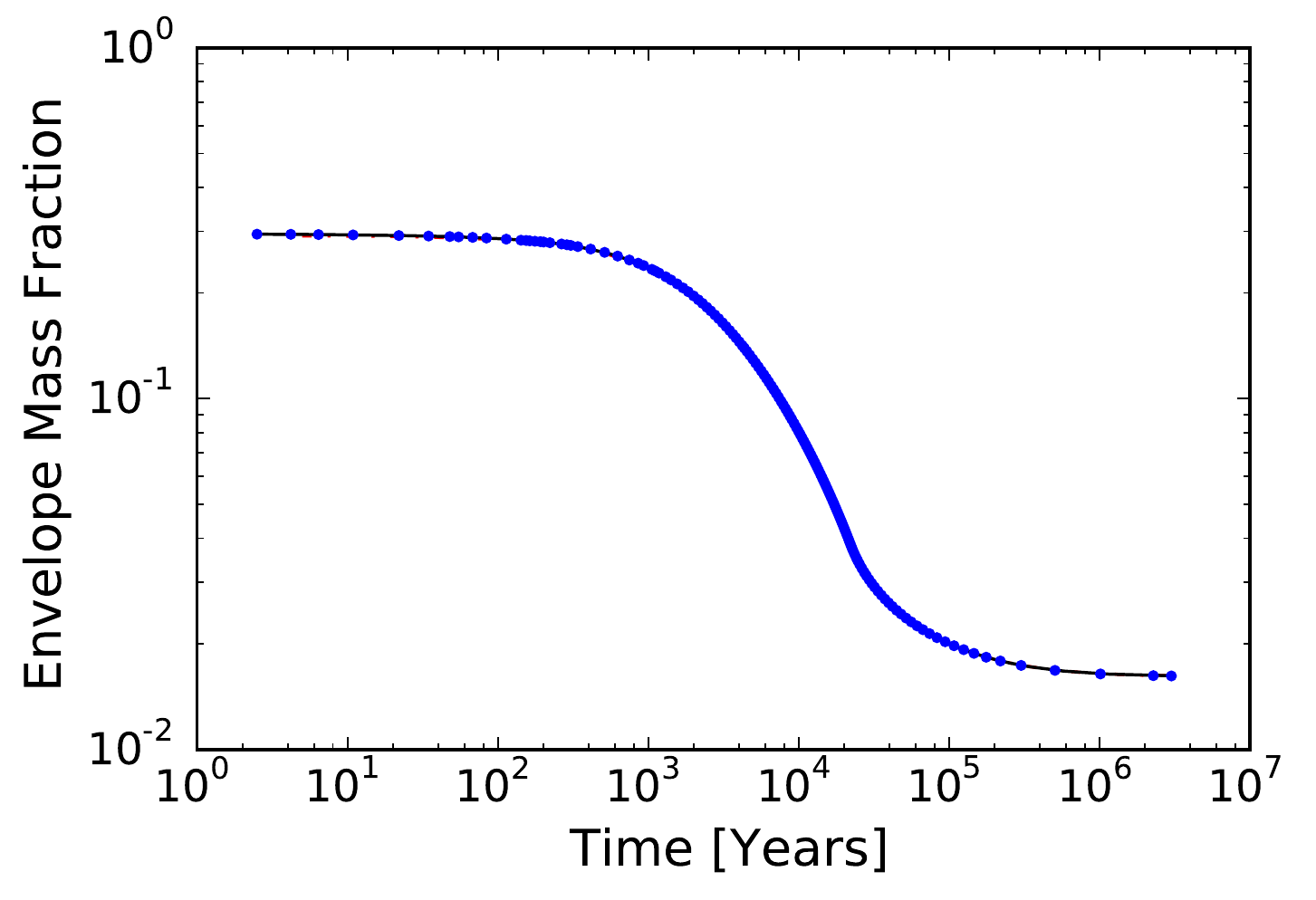}
\caption{Evolution of the envelope mass fraction for the model with a 5~M$_\oplus$ core and initial envelope mass fraction of $30\%$ at an equilibrium temperature of 900~K. The different lines show cases where the envelope mass fraction from which material can be removed as mass-loss is varied:  $<0.5\%$ (black-solid), $< 1\%$ (blue-points), $<2\%$ (red-dashed), the evolution is identical in all cases where the final envelope mass fractions at the end of the simulation are within $<0.03\%$. Therefore, we are confident the specific range of atmosphere mass over which we remove material is not driving our results.}\label{fig:env_remove}
\end{figure}
The final envelope mass fractions at the end of the simulation are within $<0.03\%$.

In our simulations we use a very simple technique to include the irradiation from the central star in the upper layers (but below the photosphere) of the planet's atmosphere. Where we heat a fixed column density of the atmosphere at the rate prescribed by stellar heating, the $F_*-\Sigma$ approach (Paxton et al. 2013). We chose to heat a column of $250$ g~cm$^{-2}$ in all our above calculations, motivated by the opacity of $4\times10^{-3}$ suggested by \citep{Guillot2010}. We can check how sensitive our results are to this choice of column density by repeating the calculation for a model with a 5~M$_\oplus$ core and initial envelope mass fraction of $30\%$. We choose to heat a column density of $50$ \& 1250 g~cm$^{-2}$ at the surface of the atmosphere, corresponding to a factor 5 change up and down of the atmospheric opacity. The results of this experiment are shown in Figure~\ref{fig:opacity_change}.
\begin{figure}
\centering
\includegraphics[width=0.5\textwidth]{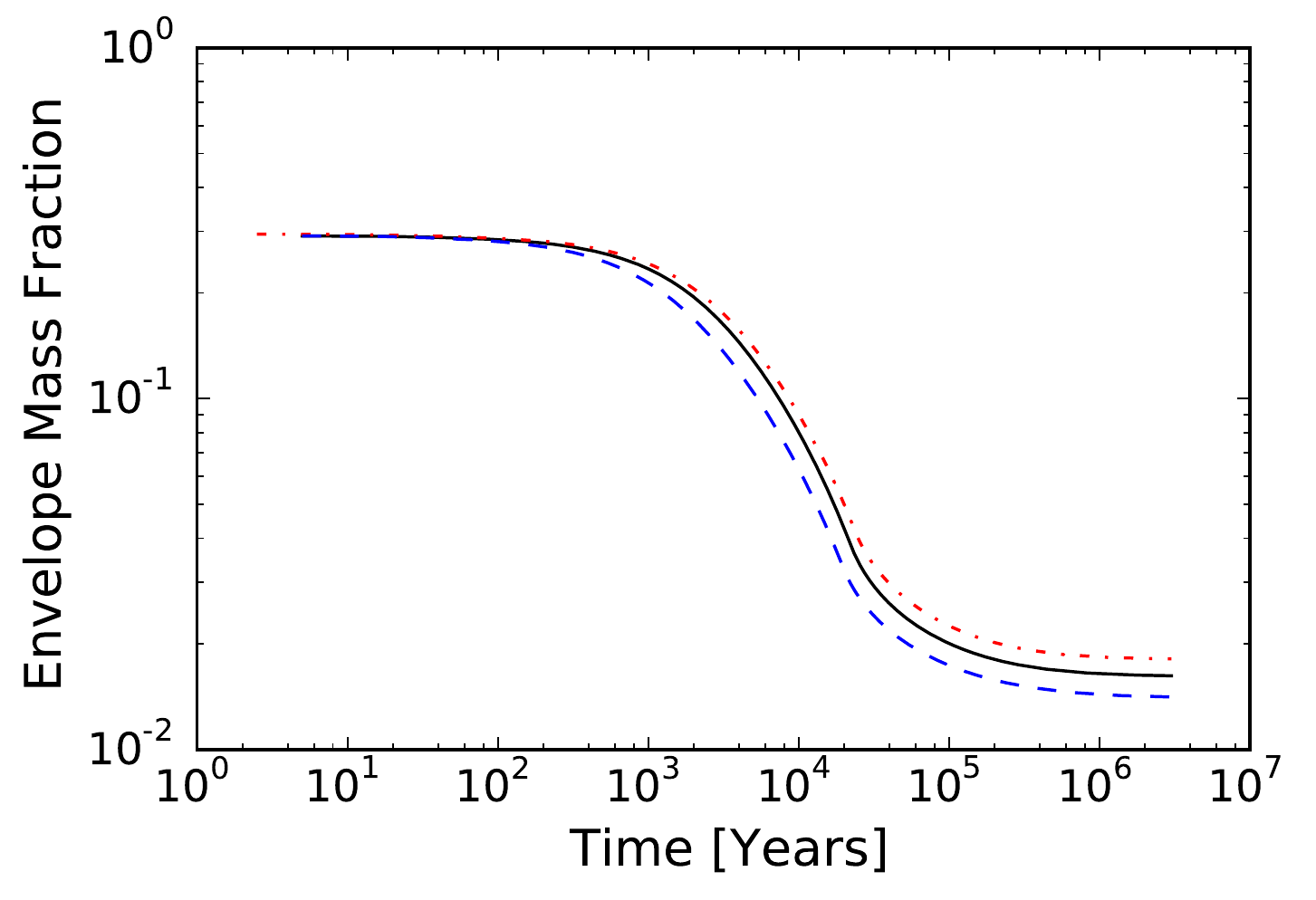}
\caption{Evolution of the envelope mass fraction for the model with a 5~M$_\oplus$ core and initial envelope mass fraction of $30\%$ at an equilibrium temperature of 900~K. The lines correspond to different column densities that are heated by the stellar irradiation, the values shown are: 50 (dashed), 250 (solid - the standard value used in our calculations) and 1250 (dot-dashed) g cm$^{-2}$.}\label{fig:opacity_change}
\end{figure}
As expected the results are obviously sensitive to the depth of the heating (as it in turn changes the radius of the planet), the effect of the column density heated by the stellar irradiation is a secondary effect and all models follow a similar evolution. 

Considering all the tests above we are happy that our model simplifications are not driving our results, which are robust.

\end{document}